\documentclass[pdflatex,iicol,sn-basic,Numbered]{sn-jnl}

\usepackage{tikz}
\usetikzlibrary{shapes.geometric, arrows.meta, positioning, fit, backgrounds,
                calc, decorations.markings}
\usepackage{pgfplots}
\pgfplotsset{compat=1.18}
\usepackage{booktabs}
\usepackage{tabularx}
\usepackage{multirow}
\usepackage{graphicx}
\hypersetup{
  hidelinks,
  pdftitle={The Stochastic Deputy: Structural Tenant Isolation for Tool-Using LLM Agents},
  pdfauthor={Mirza Samad Ahmed Baig; Syeda Anshrah Gillani; Asher Ali; Muhammad Hamzah Siddiqui},
  pdfsubject={Tenant isolation for LLM agents and Model Context Protocol servers},
  pdfkeywords={access control, large language models, Model Context Protocol, multi-tenancy, prompt injection}}
\usepackage{xcolor}
\usepackage{amsmath}
\usepackage{amssymb}
\usepackage{listings}
\usepackage{siunitx}
\definecolor{cBlue}{HTML}{2563EB}
\definecolor{cRed}{HTML}{DC2626}
\definecolor{cGreen}{HTML}{16A34A}
\definecolor{cOrange}{HTML}{EA580C}
\definecolor{cPurple}{HTML}{7C3AED}
\definecolor{cGray}{HTML}{6B7280}
\definecolor{cLightBg}{HTML}{F8FAFC}
\definecolor{cDarkBg}{HTML}{1E293B}

\newtheorem{definition}{Definition}
\newtheorem{property}{Property}
\newtheorem{assumption}{Assumption}

\begin{document}

\title[The Stochastic Deputy]{The Stochastic Deputy: Structural Tenant Isolation
for Tool-Using LLM Agents}

\author*[1]{\fnm{Mirza Samad Ahmed} \sur{Baig}}
\email{mirzasamadcontact@gmail.com}

\author[2]{\fnm{Syeda Anshrah} \sur{Gillani}}
\author[1]{\fnm{Asher} \sur{Ali}}
\author[1]{\fnm{Muhammad Hamzah} \sur{Siddiqui}}

\affil*[1]{\orgname{Fandaqah},
  \orgaddress{\city{Al Khobar}, \country{Saudi Arabia}}}
\affil[2]{\orgname{Heidelberg University},
  \orgaddress{\city{Heidelberg}, \country{Germany}}}

\abstract{
Multi-tenant tools commonly accept a tenant identifier and validate it against
the caller's entitlement. For a large language model (LLM) agent, that pattern
delegates resource selection to a process whose context may contain attacker
controlled instructions. We formalize this \emph{stochastic deputy} problem and
present a structural defense: remove tenant identity from the Model Context
Protocol (MCP) tool schema, bind scope to a verified credential, and enforce it
below the agent. In a \num{373}-trial ablation across eight model configurations
and two transports, a correctly validated tenant parameter served every
out-of-scope attempt: \num{26} of \num{26}, or \num{26} of \num{41}
plausible-pretext trials overall. With the parameter removed, no tool signature
could express the read. Twelve of \num{56} trials instead escaped the interface
by forging writable scope,
showing that interface invariance requires cryptographically protected context.
On a production dataset containing multiple GBs of data, set-valued scope caused a measured
\num{57}$\times$ latency ratio under function-wrapped membership predicates; a
\texttt{JSON\_TABLE} lateral join recovered index access where the tenant key
was indexed. The evaluation also exposes deployment limits, including an
entitlement-size query-planner cliff and incomplete index coverage. The result
is a tenant-isolation argument that depends on enforceable interfaces and
credentials rather than model compliance.
}

\keywords{
Access control, large language models, Model Context Protocol, multi-tenancy,
prompt injection.
}

\maketitle

\section{Introduction}
\label{sec:intro}

A multi-tenant service must ensure that a request issued on
behalf of one tenant cannot read another tenant's data. The overwhelmingly common
implementation of this guarantee is \emph{parameter-mediated}: the caller names
the tenant it is acting for, the service validates that the caller is entitled to
that tenant, and a filter is applied. The pattern is so routine that it is
usually invisible. A tenant-specific request carries the caller's resource
selection in a request parameter.

This works because of an assumption that is rarely stated. The caller is a
deterministic program, its behaviour is fixed by its source code, and an
adversary who wishes to change which tenant it names must first change that code
or steal its credential. Under that assumption the tenant parameter is not an
attack surface but a convenience, and validating it is sufficient.

Large language model agents violate the assumption. When an LLM invokes a tool
under the Model Context Protocol (MCP)~\cite{anthropic2024mcp} or a comparable
function-calling interface, the arguments it supplies are the output of a
stochastic process conditioned on a context window whose contents are, in any
realistic deployment, partly attacker-reachable. Prompt injection, whether direct
\cite{perez2022ignore} or indirect through retrieved content
\cite{greshake2023not}, means an adversary who can place text where the model
will read it can influence what the model asks for. An agent that summarizes
inbound email, reads a support ticket, or ingests a supplier's PDF is an agent
whose tool arguments an outsider can shape.

The consequence is precise and, we argue, under-appreciated. \emph{If the tenant
identifier is a tool parameter, the effective authority of the request is bounded
by the model's prompt rather than by the caller's credential.} The service
performs its validation faithfully. If the compromised agent's credential spans
several tenants, as is common in delegated administration, then validation passes
and the cross-tenant read succeeds. No SQL injection occurs and no control is
bypassed. The system behaves exactly as specified, and the specification is the
vulnerability.

This is a variant of Hardy's confused deputy~\cite{hardy1988confused}, in which a
privileged intermediary is induced to exercise its authority on an attacker's
behalf. What is new is that the deputy is stochastic, and the difference is not
one of degree. We state it precisely, since the rest of the paper turns on it.

\begin{definition}[Stochastic deputy]
\label{def:stochastic}
A deputy is \emph{stochastic} when all three hold:
\begin{enumerate}
    \item \textbf{Delegated authority.} It holds authority over a set of
    resources broader than the request in front of it, and selects which to
    exercise.
    \item \textbf{Non-deterministic selection.} Its selection is the output of a
    process with no guarantee of correctness, so identical inputs may yield
    different exercises of authority and no proof binds the choice to the
    requester's entitlement.
    \item \textbf{Channel confusion.} Instructions and data reach it over a
    single channel, so an attacker who can place text anywhere the deputy reads
    can compete with the principal's own instructions.
\end{enumerate}
\end{definition}

Clause 3 is what defeats the classical remedy. Confused-deputy mitigations
assume the deputy can be made to reason correctly about whose authority it
wields, whether by passing an unforgeable designator or by teaching it to check.
Both presume a deputy that can reliably distinguish an instruction from data. An
LLM cannot, not because it is poorly engineered but because that separation is
absent from its input representation. Clause 2 then removes the fallback: one
cannot certify the behaviour by testing, since passing $n$ trials bounds a rate
and not a possibility.

The consequence is that the deputy's correctness cannot be an input to the
security argument. What remains is to remove the authority from the deputy's
reach, which is what the rest of this paper does.

\subsection{Position}

Much recent work on agent security seeks to \emph{detect} injection, through
classifiers over inputs, spotlighting, or instruction hierarchies. Detection is
valuable and we do not argue against it, but it is a probabilistic defense
against a probabilistic failure, and it leaves a residual breach risk
proportional to a false-negative rate.

We take the complementary position, which is architectural rather than
behavioural:

\begin{quote}
\emph{Do not attempt to make the agent ask only for what it is entitled to. Make
the question unaskable, and place the answer's boundary where the agent cannot
reach it.}
\end{quote}

Concretely: remove the tenant identifier from the tool signature so that no token
sequence the model can emit selects a tenant; derive the scope instead from a
cryptographically bound credential; and enforce the resulting boundary at the
lowest layer the deployment permits. The security of the result is then
independent of what the model was persuaded to want. This applies Saltzer and
Schroeder's principles of complete mediation and least privilege
\cite{saltzer1975protection} at the level of interface design, and it is the
direction NIST SP~800-207 \cite{nist2020zerotrust} points for non-human
principals.

Doing this turns out to be harder than the argument suggests, for a reason that
has nothing to do with language models. Deriving the scope from a credential
rather than an argument means the scope arrives as a \emph{set}, and a set-valued
predicate is not sargable under any encoding that keeps the set out of
application hands. Readers who care about multi-tenant data access but not about
agents may find Section~\ref{sec:sarg} the most directly useful part of this
paper; it stands alone and we have not seen the result stated elsewhere.

\subsection{Contributions}

\begin{enumerate}
    \item \textbf{Problem characterization.} We formalize parameter-mediated
    authorization under a non-deterministic caller, define the stochastic deputy,
    and state a tenant-confinement goal independent of caller correctness.

    \item \textbf{Structural design and argument.} We give five invariants and
    two enforcement points, then state the assumptions and failure modes of the
    resulting security argument (Sections~\ref{sec:design} to \ref{sec:guarantee}).

    \item \textbf{Database results.} We identify transitive scope closure as a
    necessary invariant and show that set-valued scope can defeat index access.
    A lateral join restores indexed lookup when the tenant key is indexed; the
    deployed alternative exhibits an entitlement-size planner cliff
    (Sections~\ref{sec:sarg} to \ref{sec:cliff}).

    \item \textbf{An interface ablation across eight model configurations and two
    transports.} A \num{373}-trial, three-arm experiment isolates the tool
    signature, measures both plausible and injected requests, and reveals a
    bypass when agents can rewrite scope (Section~\ref{sec:ablation}).

    \item \textbf{Production-scale evaluation.} Structural, performance, and
    fail-closed tests on an operational dataset containing multiple GBs of data report both
    supporting and adverse findings, including index gaps, nullable tenant keys,
    and the weaker enforcement mode used by the measured deployment.
\end{enumerate}

\section{Background and Problem Statement}
\label{sec:background}

\subsection{Tool-calling agents}

Under MCP~\cite{anthropic2024mcp}, a server advertises a set of tools, each with
a name and a JSON-schema signature. A client, typically an LLM application,
selects a tool and emits arguments conforming to the schema. The server executes
and returns a result, which re-enters the model's context.

The security-relevant property is that \emph{the argument values are model
output}. They are not chosen by the user directly, not validated by any
deterministic intermediary, and not derivable from the user's identity. They are
sampled from a distribution conditioned on everything in the context window,
including any content the agent has retrieved.

\subsection{Parameter-mediated authorization and its assumption}

\begin{definition}[Parameter-mediated authorization]
\label{def:pma}
A service employs parameter-mediated authorization when, for a request from
principal $p$ carrying an argument $t$ naming a tenant, it returns the rows of
tenant $t$ if $t \in \mathcal{E}(p)$ and denies otherwise, where $\mathcal{E}(p)$
is the set of tenants $p$ is entitled to.
\end{definition}

The pattern's soundness rests on an unstated premise:

\begin{quote}
\textbf{(P)} For a fixed principal $p$, the value $t$ is determined by the
caller's program text, and an adversary who does not control that text or $p$'s
credential cannot influence $t$.
\end{quote}

For a conventional client, (P) holds. For an LLM client it fails, because $t$ is
a function of the context window and the context window is partly
adversary-reachable. Under that failure, the guarantee degrades from ``$p$ reads
only what $p$ is entitled to and asked for'' to:

\begin{equation}
\label{eq:degraded}
\mathrm{Reachable}(p) \;=\; \bigcup_{t \in \mathcal{E}(p)} \mathrm{Rows}(t)
\end{equation}

That is, the blast radius of a single successful injection is the caller's
\emph{entire} entitlement, not the tenant the user was working in. Equation
\eqref{eq:degraded} is the crux of the paper. Every design decision that follows
is an attempt to make $\mathrm{Reachable}(p)$ small and, critically, to make it
insensitive to model output.

In the system we study, most principals have a small scope but portfolio
operators have materially larger entitlements. The tail is what matters: a
compromised agent acting for a portfolio operator reaches every property in
that portfolio.

\subsection{Why detection is the wrong layer}

One could attempt to preserve (P) by filtering the model's inputs or constraining
its outputs. Both are worth doing and neither is sufficient here, for a reason
specific to this threat: the attack and the defense share a channel. An
instruction to the model to ``only ever request hotel $t_a$'' is itself text in the
context window, competing with attacker text on equal footing. There is no
privileged band. Detection therefore yields a probability, and multiplying a
per-request false-negative rate by a production request volume yields an expected
number of cross-tenant disclosures that is not zero.

By contrast, if no argument selects a tenant, then the set of tenant selections
expressible by the model is empty, and the probability is not small but
undefined, because the event has no representation in the interface. This is the
distinction between a stochastic control and a structural one, and it is why we
pursue the latter.

\section{Threat Model}
\label{sec:threat}

\subsection{System model}

A set of tenants $T$. A set of principals $P$, each with an entitlement
$\mathcal{E}: P \rightarrow 2^{T}$ maintained in an identity provider. A tool
server exposing tools $\mathcal{T}$ over MCP, backed by a relational database
$D$ in which most tables carry a tenant-discriminating column. An LLM agent acts
on behalf of exactly one authenticated principal per session.

\subsection{Adversary}

The adversary $\mathcal{A}$ may:

\begin{enumerate}
    \item[\textbf{A1}] \textbf{Control model context.} Insert arbitrary text into
    the agent's context window, directly as the user or indirectly via documents,
    email, web content, or database rows the agent retrieves. This subsumes all
    prompt-injection variants and grants $\mathcal{A}$ full control over which
    tools are called and with what arguments.
    \item[\textbf{A2}] \textbf{Craft transport requests.} Issue arbitrary HTTP
    requests to the tool server, including forged or replayed headers.
    \item[\textbf{A3}] \textbf{Hold a valid low-privilege credential.} Authenticate
    as some $p$ with a nonempty but restricted $\mathcal{E}(p)$.
    \item[\textbf{A4}] \textbf{Observe outputs.} Read everything returned to the
    agent.
\end{enumerate}

The adversary may \emph{not} forge signatures under the identity provider's
private key, obtain database credentials, execute code on the server, or modify
server source. We assume standard transport security.

A1 is deliberately maximal, and it is the modelling counterpart of
Definition~\ref{def:stochastic}: granting the adversary the agent's behaviour
outright is how one reasons about a deputy whose selection is
non-deterministic and whose instruction channel the adversary shares. We do not
model the LLM as partially trustworthy or as resisting injection with some
probability. We grant $\mathcal{A}$ complete
control of the agent's behaviour and require the security goal to hold
regardless. This is what makes the resulting guarantee independent of model
quality, and it is why we consider an empirical injection-success study
orthogonal rather than necessary: a defense that survives total agent compromise
is not made stronger by evidence that compromise is difficult.

\subsection{Security goal}

\begin{definition}[Scope confinement]
\label{def:goal}
For every principal $p$ and every sequence of tool invocations $\sigma$ issued in
$p$'s session, including sequences chosen adversarially under A1, the set of
rows returned satisfies
\[
\mathrm{Rows}(\sigma) \subseteq \bigcup_{t \in \mathcal{E}(p)} \mathrm{Rows}(t).
\]
\end{definition}

Scope confinement is the strongest goal achievable without also constraining
\emph{which} of $p$'s tenants the agent may act for within a session, a question
we address separately through intra-tenant role enforcement
(Section~\ref{sec:i4}) but do not fold into the primary goal.

Table~\ref{tab:threats} enumerates the attack vectors this goal must survive and
names, for each, the layer that has to fail before it succeeds.

\begin{table}[!t]
\centering
\caption{Attack vectors and the invariant that addresses each. ``Enforced at''
names the layer that must fail for the attack to succeed.}
\label{tab:threats}
\renewcommand{\arraystretch}{1.25}
\begin{tabularx}{\linewidth}{@{} >{\raggedright\arraybackslash}p{1.35cm} X >{\raggedright\arraybackslash}p{1.5cm} @{}}
\toprule
\textbf{Vector} & \textbf{Description} & \textbf{Enforced at} \\
\midrule
V1: Argument \newline steering & Agent emits a tenant id it is not entitled to & Interface (I1): no such argument \\
\addlinespace
V2: Header \newline injection & Client asserts identity via HTTP header & Gateway (I2) \\
\addlinespace
V3: Token \newline forgery & Claims modified in transit & JWKS signature (I2) \\
\addlinespace
V4: Forgotten \newline predicate & New tool omits the tenant filter & Choke point or engine (I3) \\
\addlinespace
V5: Join \newline traversal & Scoped primary joined to unscoped table & Builder (I4) \\
\addlinespace
V6: Role \newline escalation & Low-role token invokes sensitive tool & Pre-query check (I4) \\
\addlinespace
V7: Excess \newline disclosure & Tool returns PII beyond purpose & Egress boundary (I5) \\
\bottomrule
\end{tabularx}
\end{table}

\begin{figure*}[!t]
\centering
\includegraphics[width=\textwidth]{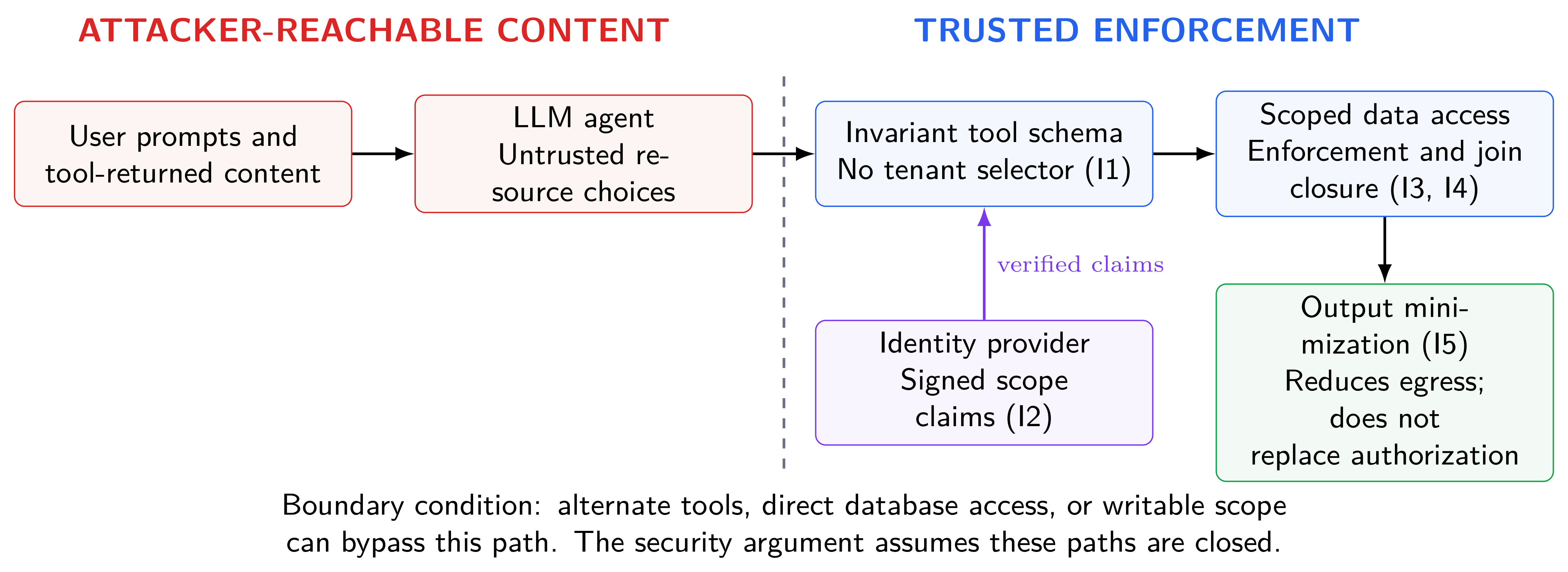}
\caption{Trust-boundary map. Attacker-reachable content can influence the agent,
but must not determine the verified scope. Dashed separation marks the intended
boundary, not a guarantee against capabilities that bypass it.}
\label{fig:trustmap}
\end{figure*}

Figure~\ref{fig:trustmap} summarizes the trust boundary assumed by the
design. Agent-influenced content must not determine verified scope or
open an alternative data-access path.

\section{Design}
\label{sec:design}

The architecture rests on five invariants. We state each as a property of the
system rather than a component, because the point is what cannot happen, not what
is present.

A word on what ``invariant'' claims here, since the term is sometimes read as
promising more than we deliver. These are properties the implementation is built
to maintain, and Section~\ref{sec:guarantee} argues they compose into the
isolation goal. That argument is a conditional one, relative to four assumptions
we state explicitly and then attack. It is not a machine-checked proof: we have
no formal model of the system, no mechanized refinement from specification to
code, and consequently no guarantee that the implementation realizes the
invariants everywhere. What we offer instead is empirical: a static analysis
establishing that every tool reaches the database through the one path that
enforces them, and an operational-principal test that the composition holds end
to end for an anonymized entitlement, both in Section~\ref{sec:realprincipal}. A reader who
wants I1 through I5 as theorems
should treat this paper as motivating that work rather than as having done it.

\begin{figure*}[!t]
\centering
\includegraphics[width=\textwidth]{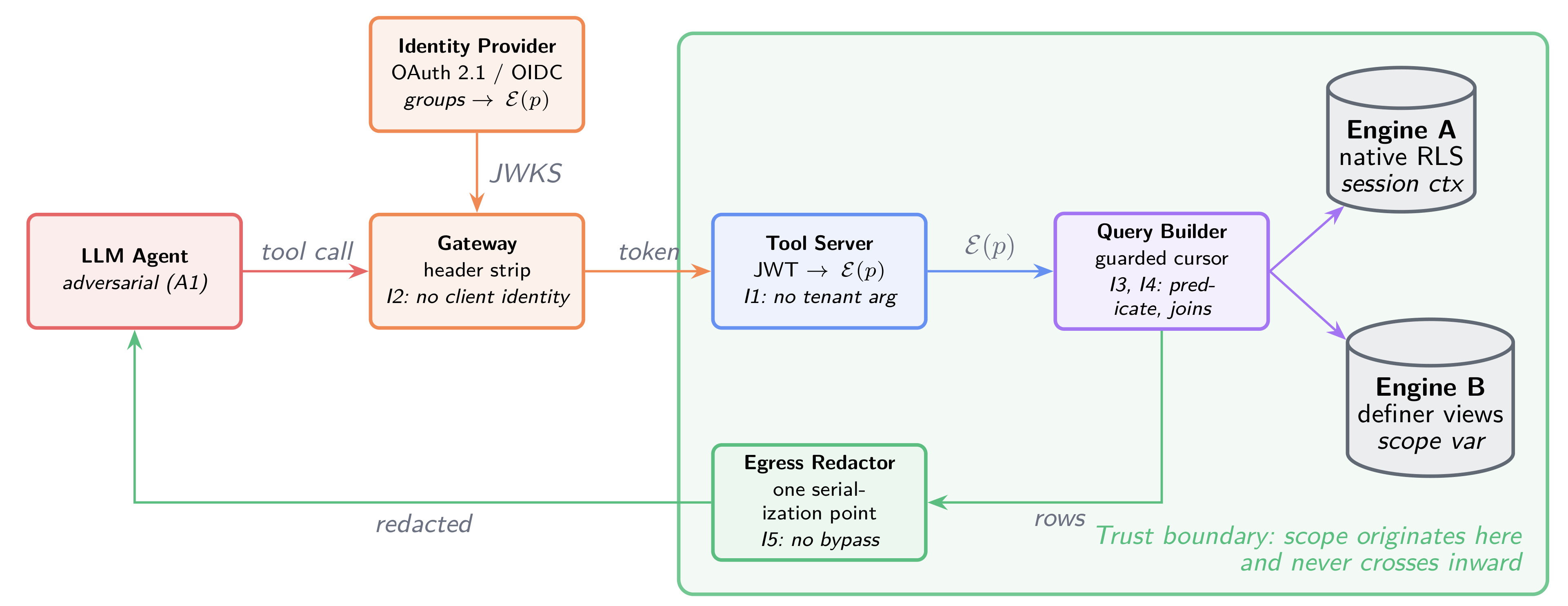}
\caption{Architecture. The tenant scope enters at the identity provider and flows
\emph{downward} only. No path exists by which agent output influences the scope,
because no tool schema accepts a tenant argument (I1) and no client-supplied
header survives the gateway (I2). The two database engines illustrate the
enforcement-point design space of Section~\ref{sec:i3}.}
\label{fig:arch}
\end{figure*}

Fig.~\ref{fig:arch} shows how the five compose. The tenant scope enters at the
identity provider and travels downward only; no arrow runs from the agent to
anything that determines scope, which is the property the rest of this section
establishes piece by piece.

\subsection{I1: Interface invariance}
\label{sec:i1}

\begin{property}[Interface invariance]
No tool schema in $\mathcal{T}$ admits an argument whose value selects a tenant.
\end{property}

This is the load-bearing invariant and the cheapest to implement. A tool is
\texttt{get\_reservations(status, date\_from, limit)}, not
\texttt{get\_reservations(hotel\_id, \ldots)}. Where a tool must operate on one
of several entitled tenants, the tenant is selected by a preceding tool that
returns only $\mathcal{E}(p)$, and the selection is validated against
$\mathcal{E}(p)$ server-side, so the argument is a choice \emph{within} the
entitlement and never a widening of it.

The security content is a statement about expressiveness. The language of
well-formed tool calls contains no sentence that requests another tenant's data.
Under A1 the adversary controls the model completely and this does not help,
because the constraint is on the schema, not on the sampling.

\subsection{I2: Cryptographic context binding}
\label{sec:i2}

\begin{property}[Context binding]
$\mathcal{E}(p)$ is derived solely from claims in a token whose signature is
verified against the identity provider's published JWKS, and never from any
other part of the request.
\end{property}

The reverse proxy strips client-supplied identity headers unconditionally, so a
header asserting identity cannot reach application code even if application code
would have honoured it (V2). Entitlements travel as group claims of the form
\texttt{/tenants/\textless id\textgreater/\textless role\textgreater}.

Claim parsing is a security boundary and is treated as one. Two subtleties are
worth recording because both were live defects found during development.

\emph{Unicode digit conflation.} Parsing the tenant id with the regular
expression class \verb|\d| is unsafe in Python, where \verb|\d| matches every
Unicode decimal digit. A group path written with Arabic-Indic or Devanagari
digits then passes validation, and \texttt{int()} folds it to the same integer as
its ASCII spelling, so several visually distinct group names, which an
administrator reviewing the identity provider console would read as distinct,
collapse onto one tenant. The parser restricts the class to \verb|[0-9]| and
additionally requires canonical form with no leading zeros, rejecting
\texttt{/tenants/0$t_a$/manager}.

\emph{Refuse versus skip.} A malformed grant must be refused, not skipped. A
skipped grant silently narrows a user's access and presents as a functional
fault rather than a security event, so it is diagnosed slowly and wrongly. The
one exception is a path under the tenant prefix that does not claim to be a grant
at all. An identity provider renders the group tree such that joining an
intermediate node instead of a leaf is a single misclick, and that case is
skipped and recorded, because refusing the entire token for it locked a real
account out of tenants it legitimately held.

\subsection{I3: Enforcement relocation, and its design space}
\label{sec:i3}

\begin{property}[Enforcement relocation]
The tenant predicate is applied at a layer below the tool implementation, such
that a tool cannot emit a query lacking it.
\end{property}

Application-layer filtering distributed across tools fails by omission. With $n$
tools over $m$ tenant-scoped tables there are $O(nm)$ sites at which a predicate
can be forgotten, each omission produces a silent cross-tenant read, and none
produces an error.

Where the predicate should instead live is not a single answer but a choice
constrained by how much authority one has over the database. We characterize two
points, because reporting only the stronger one, as the enforcement-relocation
literature tends to, misrepresents what most teams can actually deploy.

\subsubsection{Point A: engine-enforced (definer-rights views)}

The application account is granted \texttt{SELECT} on a set of generated views
and on \emph{no base table}. Each view joins its base table against a
definer-rights function returning the session scope. Because the view runs with
the definer's privileges and the application account has none of its own, a query
naming a base table fails with \texttt{ERROR 1142} rather than returning
unfiltered rows.

The guarantee is that a bypass \emph{fails closed at the engine}. The cost is a
generated view per tenant-scoped relation, helper functions, and least-privilege
accounts on the primary, which is a production database change requiring
change-control approval. On engines with native row-level security (RLS), this
point is reached directly by an RLS policy and a session-context assignment, with
the additional benefit, absent in the view construction, that the session context
can be marked read-only for the connection's lifetime.

\subsubsection{Point B: application choke point}

Where the database cannot be altered, the predicate lives in a single
application module through which all queries must pass. Two mechanisms make
``must'' structural rather than advisory:

\begin{itemize}
    \item The cursor accepts only a \texttt{Query} object, never a string.
    Passing raw SQL raises \texttt{TypeError} at the call site.
    \item A \texttt{Query} is obtainable only from a builder that cannot be
    constructed without an entitlement and that unconditionally appends the
    predicate. There is no unscoped mode and no disabling flag.
\end{itemize}

\begin{lstlisting}[language=Python, caption={The choke point. The type error is the enforcement: a forgotten predicate becomes a crash at the call site rather than a silent cross-tenant read.}, label={lst:guard}]
class GuardedCursor:
    def execute(self, query):
        if not isinstance(query, Query):
            raise TypeError(
                "GuardedCursor.execute takes a Query from "
                "TenantQuery, not %s. Raw SQL cannot be "
                "executed here." % type(query).__name__)
        self._audit_tables.extend(query.tables)
        self._cur.execute(query.sql, query.params)
        return self
\end{lstlisting}

The guarantee is weaker and we state it precisely. \emph{The omission failure
mode is eliminated, but the containment is not.} The database account can read
every row of every table, because nothing in the database restricts it. Any code
path that obtains a connection without going through the module bypasses
everything. Point B converts a silent leak into a loud crash; it does not convert
a bypass into a denial. Table~\ref{tab:modes} sets the two points side by side;
the row that matters is the last one, which states what a bypass costs.

\begin{table}[!t]
\centering
\caption{The two enforcement points. The distinction that matters is the last
row: what a bypass yields.}
\label{tab:modes}
\renewcommand{\arraystretch}{1.25}
\begin{tabularx}{\linewidth}{@{} >{\raggedright\arraybackslash}p{2.05cm} X X @{}}
\toprule
 & \textbf{A: engine} & \textbf{B: choke point} \\
\midrule
Filter resides in & views or RLS policy & application module \\
\addlinespace
DB change & per-table views and helpers & none \\
\addlinespace
Forgotten predicate & impossible & \texttt{TypeError} \\
\addlinespace
Scope tamperable & no (read-only context, engine A) & yes (session variable) \\
\addlinespace
Retains indexed plan beyond builder flip & yes & no (Sec.~\ref{sec:cliff}) \\
\addlinespace
\textbf{Bypass yields} & \textbf{\texttt{ERROR 1142}} & \textbf{every tenant} \\
\bottomrule
\end{tabularx}
\end{table}

Because the two are not equally safe, the implementation refuses to start unless
the operator names one explicitly. There is no default. A default would be a
security decision made by whoever wrote the configuration parser.

\subsection{I4: Transitive scope closure}
\label{sec:closure}
\label{sec:i4}

\begin{property}[Transitive closure]
Every tenant-owned relation reachable in a query, not only the primary, carries
the tenant predicate.
\end{property}

This invariant is, in our experience, the one most often missed, and it is the
one with the largest measured exposure. A predicate on the primary table does not
constrain a joined table. The query
\begin{center}
\texttt{FROM reservations r LEFT JOIN customer c ON c.id = r.customer\_id}
\end{center}
is correctly scoped on \texttt{reservations} and entirely unscoped on
\texttt{customer}.

Whether this leaks depends on whether the foreign key ever crosses a tenant
boundary. In the studied system it does because a shared registry can associate
a returning guest's booking at tenant B with a record created by tenant A. The
highest-rate audited relationship crosses in \SI{15.72}{\percent} of rows, and
the crossing records carry direct identifiers (Section~\ref{sec:rq4}). The
naive join therefore discloses guest names and contact details across a tenant
boundary at scale.

The design response is to make a join a declared object rather than a SQL
fragment. The builder then, without the tool author's participation, refuses
tables outside the allowlist, appends the tenant equality between the joined and
primary relations, appends the soft-delete filter where the joined table has one,
rewrites to the view form under enforcement point A, and records the joined table
in the audit line. That last item is not incidental. When joins were free
strings, the audit record named only the primary table, so a join reading
\texttt{customer} was simultaneously unscoped and invisible to the mechanism
meant to detect it afterwards.

Intra-tenant role enforcement is applied at the same layer. The role component of
the group claim is checked before query construction, so a low-privilege token
invoking a sensitive tool is denied without a database round trip (V6).

\subsection{I5: Egress-boundary minimization}
\label{sec:i5}

\begin{property}[Egress minimization]
Personal data is redacted at the single serialization function through which all
tool results pass, unconditionally.
\end{property}

Per-tool redaction is redaction some future tool will omit. With several dozen
tools, the next one is the problem. Placing the pass at the serialization boundary
means a newly written tool that selects an email column is redacted before anyone
notices it was exposed. The trade is that redaction keys on column name rather
than on a semantic type, which biases toward over-masking. A switchboard number
in a column named \texttt{phone} is masked too. We consider that the correct
direction for the bias.

Two design decisions warrant defense. First, guest \emph{names} are not masked.
Purpose limitation under Saudi PDPL~\cite{ksa2023pdpl} and
GDPR~\cite{eu2016gdpr} requires processing to be necessary for the purpose, not
that every identifier be removed, and a front-desk manager asking who is arriving
today cannot be served by a redacted list. Contact details and dates of birth are
not necessary for that purpose and are masked. Second, there is no disabling
switch, no environment variable, no role and no argument, and the test suite
asserts that plausible ones have no effect. A switch that turns compliance off is
a switch that will be found turned off during an incident.

\begin{figure*}[!t]
\centering
\includegraphics[width=\textwidth]{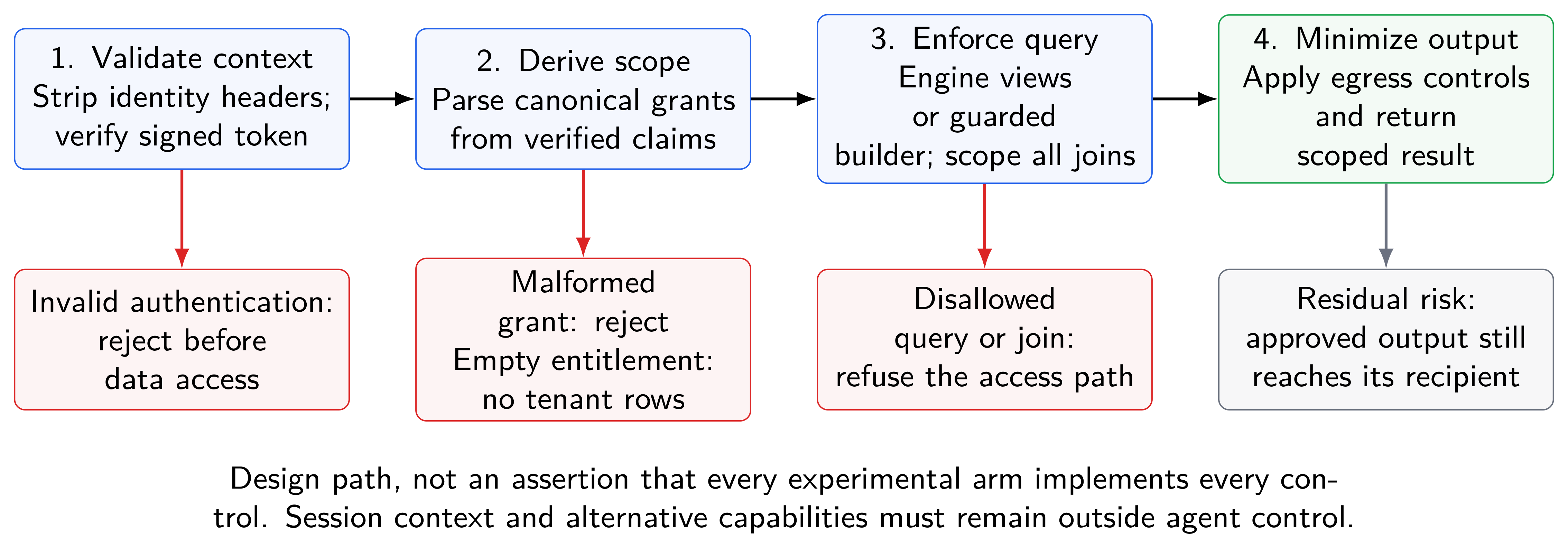}
\caption{Conceptual request flow and rejection paths for the structural design.
Engine enforcement and the guarded builder are alternative enforcement points.
Output minimization reduces exposure but does not eliminate authorized egress.}
\label{fig:requestflow}
\end{figure*}

Figure~\ref{fig:requestflow} summarizes the enforcement sequence and its
rejection paths. It depicts the structural design, not additional
experimental results.

\subsection{Security argument and its assumptions}
\label{sec:guarantee}

We now argue Definition~\ref{def:goal} holds and, more usefully, state what each
step depends on.

Fix a principal $p$ and an adversarial invocation sequence $\sigma$ under A1.

\emph{Step 1.} By I1, no argument in any call in $\sigma$ names a tenant.
Therefore the tenant scope used to serve $\sigma$ is not a function of $\sigma$.

\emph{Step 2.} By I2, the scope is a function of the verified token alone, hence
equals $\mathcal{E}(p)$. A2 does not help, because headers are stripped. A3 gives
$\mathcal{A}$ a token, but for a principal whose entitlement is by definition
permitted.

\emph{Step 3.} By I3, every emitted query carries the predicate
$\text{tenant} \in \mathcal{E}(p)$; by I4 this extends to joined relations. Hence
every returned row belongs to some $t \in \mathcal{E}(p)$, which is
Definition~\ref{def:goal}.

The argument is only as strong as four assumptions, which we state so they can be
attacked.

\begin{assumption}[Complete mediation]
\label{as:mediation}
All data egress passes through the builder and the serialization boundary.
\end{assumption}
Violation is the dominant residual risk under enforcement point B, where a
connection obtained outside the module reads everything. Under point A, violation
degrades to \texttt{ERROR 1142}. This asymmetry \emph{is} the argument for
point~A.

\begin{assumption}[Correct entitlement mapping]
\label{as:mapping}
The identity provider's group claims reflect intended entitlements.
\end{assumption}
We enforce the syntax of claims, not their truth. An administrator who grants the
wrong tenant produces a correct system serving a wrong policy. This is outside
the threat model but inside the risk.

\begin{assumption}[Scope integrity for the connection's lifetime]
\label{as:integrity}
The stamped scope is not modified after being set.
\end{assumption}
Holds by construction on engines offering a read-only session context. It does
\emph{not} hold on MySQL, whose user variables have no read-only mode: the
application account may reassign the scope variable at any time. The risk is
reduced from every query in every tool to one line of connection setup plus any
injection reaching that connection, which is a real improvement and not parity
with RLS.

\begin{assumption}[No session-state reuse]
\label{as:pooling}
A connection carrying one principal's scope is never reused for another.
\end{assumption}
Enforced by using connection-per-request and no pooling. A pooled connection
carries its scope variable to whoever checks it out next, a cross-tenant read
with no incorrect SQL anywhere and nothing in the audit trail to show it. This is
the least obvious of the four and, we suspect, the most likely to be
reintroduced by a well-meaning performance optimization. It is also not free:
Section~\ref{sec:conncost} measures its cost, which is substantially larger than
we assumed when adopting it.

\section{Set-Valued Scope and Index Sargability}
\label{sec:sarg}

We now turn to an obstacle that any enforcement-relocation design meets, that is
independent of LLMs and of our architecture, and that we have not seen reported.

\subsection{Why the set is the hard case}

Textbook RLS assumes a scalar scope. The session carries one tenant, and the
predicate is an equality the optimizer resolves as an index seek. Real
entitlements are sets: a portfolio manager may hold many properties.

A set-valued scope must be transported into the query somehow, and the transport
turns out to determine whether the query is $O(\log n)$ or $O(n)$. The scope
cannot be inlined as a literal \texttt{IN} list, because the whole point of
enforcement relocation is that the predicate is not assembled per-call by
application code; it must come from a session variable or a function that a view
can reference. And the natural ways to test membership of a column against a
session-held collection are all \emph{function-wrapped}: the column appears as an
argument to a function rather than bare on one side of a comparison. A
function-wrapped column is not sargable, so the optimizer cannot use the index
\cite{selinger1979access}, and the predicate degenerates to a scan.

This is a genuine tension, not an implementation slip. The constructions that
make the predicate tamper-resistant are precisely the ones that hide the column
from the optimizer.

\subsection{Encodings}

Let $S$ be the session scope. Four candidate encodings:

\begin{enumerate}
    \item[E1] \texttt{JSON\_CONTAINS(@s, CAST(team\_id AS JSON))}
    \item[E2] \texttt{team\_id MEMBER OF (CAST(@s AS JSON))}
    \item[E3] \texttt{FIND\_IN\_SET(team\_id, @csv)}
    \item[E4] \texttt{JOIN JSON\_TABLE(@s, '\$[*]' COLUMNS (t INT PATH '\$')) jt\\ \hspace*{1em} ON t.team\_id = jt.t}
\end{enumerate}

E1 to E3 wrap the column. E4 does not. It materializes the scope as a derived
relation of $|S|$ rows and joins, leaving \texttt{team\_id} bare on one side of an
equality. The optimizer drives from the small derived table and performs one
index \texttt{ref} lookup per scope element. The rewrite is semantically
equivalent, since a membership test is a semijoin, but operationally it is the
difference between a seek and a scan.

A fifth form, the inline literal \texttt{IN} list, is sargable and serves as our
performance reference. It is not, however, a candidate encoding: assembling it
requires the application to interpolate the scope into each statement, which is
exactly the arrangement enforcement relocation exists to remove. We measure it to
establish what a seek costs, not as an option.

\subsection{Measured effect}
\label{sec:sargmeasured}

Table~\ref{tab:sarg} reports \texttt{EXPLAIN FORMAT=JSON} and interleaved latency
measurements on the largest indexed operational table under a synthetic
three-tenant scope. The table name and absolute cardinality are withheld;
optimizer work is normalized to the full-index-scan estimate. Every encoding
returned an identical result.

\begin{table}[!t]
\centering
\caption{Access path and normalized optimizer work. Latency is the median of
round-robin interleaved trials (Section~\ref{sec:latencymethod}).}
\label{tab:sarg}
\renewcommand{\arraystretch}{1.22}
\begin{tabularx}{\linewidth}{@{} X l r r @{}}
\toprule
\textbf{Encoding} & \textbf{Access} & \textbf{Rel. work} & \textbf{Med. ms} \\
\midrule
E1 \texttt{JSON\_CONTAINS} & \texttt{index} & \SI{100}{\percent} & \num{5781.24} \\
E3 \texttt{FIND\_IN\_SET}  & \texttt{index} & \SI{100}{\percent} & \num{1330.02} \\
Inline \texttt{IN}         & \texttt{range} & \SI{6.25}{\percent} & \num{81.36} \\
\textbf{E4 \texttt{JSON\_TABLE}} & \textbf{\texttt{ref}} & \textbf{\SI{0.085}{\percent}} & \num{101.36} \\
Deployed builder (point B) & \texttt{range} & \SI{6.25}{\percent} & \num{453.31} \\
\bottomrule
\end{tabularx}
\end{table}

The function-wrapped forms select the tenant index but traverse it end to end.
E2 produces the same plan as E1. In contrast, E4 drives indexed
\texttt{ref} lookups from the derived scope table, reducing the optimizer's
estimated work by \SI{99.92}{\percent}. The access method, not the absolute
estimate, is the robust signal because estimates for \texttt{range} and
\texttt{ref} are not directly commensurable.

The latency consequence is material: E1 is \num{57}$\times$ slower than E4 in
the interleaved campaign. E4 is not faster than a hand-written \texttt{IN}
list, however. It is slower by a median of \SI{27.6}{\milli\second} and wins only
\num{9} of \num{25} paired rounds. This corrects an earlier block-timed result.
The defensible claim is that \texttt{JSON\_TABLE} restores indexed access at
comparable cost while remaining expressible inside a view.
\subsection{The result is conditional on an index}
\label{sec:sargindex}

Extending the comparison across every allowlisted tenant table yields the
conditional result in Table~\ref{tab:sargmatrix}. \texttt{JSON\_TABLE} obtains a
seek on approximately four fifths of the audited tables. On the remainder, every
encoding degrades to \texttt{ALL}; these are exactly the tables without a tenant
index.

\begin{table}[!t]
\centering
\caption{Access path across the allowlist, grouped to avoid disclosing schema
cardinality or business table names.}
\label{tab:sargmatrix}
\renewcommand{\arraystretch}{1.18}
\begin{tabularx}{\linewidth}{@{} X l l l @{}}
\toprule
\textbf{Tenant-key index} & \textbf{E4 JT} & \textbf{Inline IN} & \textbf{E1 JC} \\
\midrule
Present (approximately four fifths) & \texttt{ref} & \texttt{range} & \texttt{index} \\
Absent (approximately one fifth) & \texttt{ALL} & \texttt{ALL} & \texttt{ALL} \\
\bottomrule
\end{tabularx}
\end{table}

The rewrite does not create sargability; it stops destroying it. An adopter must
therefore audit tenant-key indexes before expecting this result to transfer.
\subsection{Heterogeneous key types}

A further failure is silent. One legacy tenant key is stored as text while the
remaining audited keys are numeric. Binding an integer scope against the text
column forces conversion and abandons the seek. Normalizing the best plan to
one unit of optimizer work gives the measured separation below.

\begin{center}
\footnotesize
\begin{tabular}{@{}llr@{}}
\toprule
\textbf{Binding} & \textbf{Access} & \textbf{Relative work} \\
\midrule
integer-bound \texttt{IN} & \texttt{index} & \num{1525}$\times$ \\
string-bound \texttt{IN} & \texttt{range} & \num{113}$\times$ \\
\texttt{JSON\_TABLE}, collation pinned & \texttt{ref} & \num{1}$\times$ \\
\bottomrule
\end{tabular}
\end{center}

The builder therefore binds that scope as strings and pins the target column's
collation in \texttt{JSON\_TABLE}; otherwise an implicit collation conversion
can recreate the scan. Tenant-key type and collation are consequently part of
the security performance contract, not incidental schema metadata.

\section{What the Deployed Mode Costs}
\label{sec:cliff}

The sargability benefit of Section~\ref{sec:sarg} accrues to enforcement point A,
not to the mode in production. Point B's builder emits an inline \texttt{IN}
predicate, because there is no view to hold the join, and adds two guards the
bare comparison lacks: the soft-delete filter and the correlated \texttt{EXISTS}
against \texttt{teams} that suppresses ghost data (Section~\ref{sec:rq4}). This
section measures what that costs and, more usefully, why.

\subsection{Overhead decomposition}
\label{sec:decomp}

Table~\ref{tab:decomp} adds the two guards one at a time to an otherwise
identical statement, warmed and interleaved, with \texttt{EXPLAIN} captured for
each variant.

\begin{table}[!t]
\centering
\caption{Overhead decomposition on \texttt{reservations}, three-tenant scope,
warmed. The soft-delete filter, not the ghost guard, is the dominant term. The
\texttt{filtered} column is the optimizer's estimate of the fraction of
index-matched rows surviving the residual predicate.}
\label{tab:decomp}
\renewcommand{\arraystretch}{1.22}
\begin{tabularx}{\linewidth}{@{} X r r c @{}}
\toprule
\textbf{Variant} & \textbf{Med. ms} & \textbf{$\Delta$ ms} & \textbf{filtered} \\
\midrule
Bare \texttt{IN} (baseline)            & \num{65.42}  & \num{0}        & \SI{100}{\percent} \\
\quad + soft-delete filter             & \num{411.73} & $+\num{346.31}$ & \SI{50}{\percent} \\
\quad + ghost guard                    & \num{449.87} & $+\num{38.14}$  & \SI{50}{\percent} \\
Ghost guard alone                      & \num{101.62} & $+\num{36.20}$  & \SI{100}{\percent} \\
\midrule
\textbf{Deployed builder}              & \textbf{\num{452.09}} & $+\num{386.67}$ & \SI{50}{\percent} \\
\bottomrule
\end{tabularx}
\end{table}

\begin{figure*}[!t]
\centering
\includegraphics[width=0.70\textwidth]{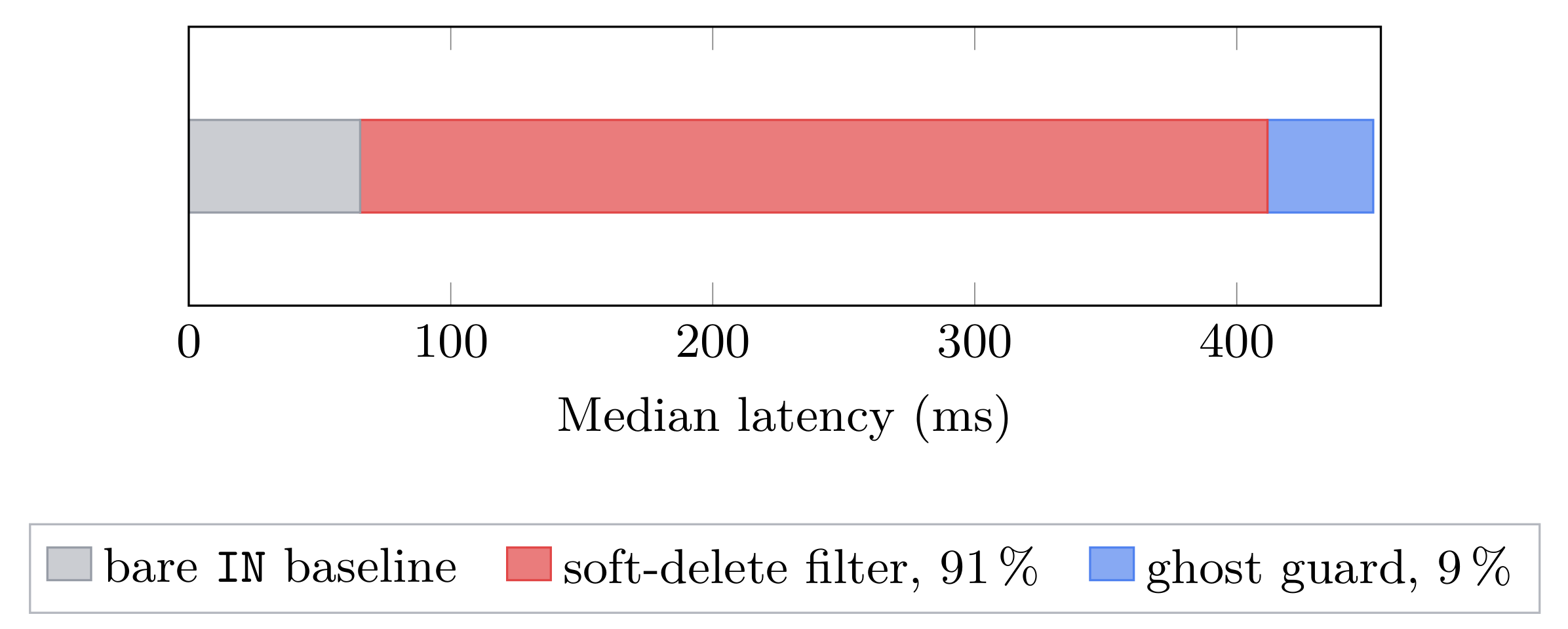}
\caption{Where the deployed mode's \SI{452}{\milli\second} actually goes. Of the
\SI{387}{\milli\second} above the bare baseline, the soft-delete filter accounts
for \SI{346}{\milli\second}, which is \SI{91}{\percent}. The ghost guard, which
an earlier version of this work blamed on the strength of its apparent
complexity, accounts for \SI{9}{\percent}. Most of what looks like the cost of
isolation is a missing composite index.}
\label{fig:decomp}
\end{figure*}

The attribution is unambiguous and it corrects an earlier claim of ours, and
Fig.~\ref{fig:decomp} shows the split at a glance. The
soft-delete filter accounts for \SI{91}{\percent} of the overhead
(\SI{346}{\milli\second} of \SI{387}{\milli\second}); the ghost guard accounts
for \SI{9}{\percent} (\SI{36}{\milli\second} measured in isolation). We had
previously attributed the cost to the ghost guard, reasoning from its apparent
complexity as a correlated subquery rather than from measurement.

The mechanism is visible in the \texttt{filtered} estimate, which drops from
\SI{100}{\percent} to \SI{50}{\percent} the moment the soft-delete filter is
added. The tenant index locates candidate rows, but \texttt{deleted\_at} is not
part of that index, so the engine must fetch and test every candidate. The
correlated \texttt{EXISTS}, by contrast,
resolves as a unique primary-key lookup per distinct tenant and is nearly free.

\textbf{The fix is an index, not an architecture change.} We audited index
coverage on the largest tenant-scoped tables and found no composite index
leading with the tenant column and including \texttt{deleted\_at}
(every one indexes the tenant column alone). Adding
\texttt{(team\_id, deleted\_at)} would make the soft-delete predicate
index-resident and should collapse the dominant term. We could not verify this
directly, because the replica is \texttt{super\_read\_only} and index creation
requires the primary, so we report it as a diagnosis rather than a demonstrated
remedy.

\subsection{A query-planner cliff at lower-double-digit scopes}
\label{sec:cliffmeasured}

The more serious finding is not constant overhead but a discontinuous plan
change. Table~\ref{tab:cliff} reports both enforcement points across synthetic
entitlement sizes. Absolute optimizer estimates are replaced by work normalized
to the post-flip scan so the table does not disclose operational cardinality.

\begin{table*}[!t]
\centering
\caption{Scaling with entitlement size. Between $|S|=10$ and $|S|=15$, the
deployed builder abandons the tenant index; the views form retains indexed
\texttt{ref} access. Blank cells were not measured.}
\label{tab:cliff}
\renewcommand{\arraystretch}{1.14}
\begin{tabularx}{0.78\textwidth}{@{} c r l r r @{}}
\toprule
& \multicolumn{3}{c}{\textbf{Builder (point B)}} & \textbf{Views (A)} \\
\cmidrule(lr){2-4}\cmidrule(lr){5-5}
$|S|$ & \textbf{Med. ms} & \textbf{Access} & \textbf{Rel. work} & \textbf{Med. ms} \\
\midrule
1  & \num{1378}  & \texttt{index\_merge} & \num{0.03} & \num{186} \\
2  & \num{339}   & \texttt{range}        & \num{0.09} & \\
3  & \num{444}   & \texttt{range}        & \num{0.13} & \num{383} \\
5  & \num{633}   & \texttt{range}        & \num{0.18} & \\
10 & \num{1065}  & \texttt{range}        & \num{0.33} & \num{892} \\
\midrule
15 & \textbf{\num{10005}} & \texttt{ref} on deletion key & \num{1.00} & \\
20 & \num{10117} & \texttt{ref} on deletion key & \num{1.00} & \num{1549} \\
30 & \num{10819} & \texttt{ref} on deletion key & \num{1.00} & \\
\bottomrule
\end{tabularx}
\end{table*}

Below the flip, point B uses the tenant index and latency grows with scope. At
$|S|=15$, the optimizer instead selects the soft-delete index and performs
approximately constant work regardless of scope, producing an order-of-magnitude
step. The singleton is a third \texttt{index\_merge} plan rather than part of
the trend.

The missing composite index explains both the small-scope overhead and the
cliff. Point A retains tenant-key \texttt{ref} access throughout the measured
range and has no discontinuity; at $|S|=20$, point B is approximately
\num{6.5}$\times$ slower. Thus, in this deployment the stronger enforcement
point is also the faster one.
\subsection{Connection-per-request cost}
\label{sec:conncost}

Assumption~\ref{as:pooling} forbids connection pooling, and we previously
described its cost as negligible for the measured workload. That description was
not based on a measurement. It is wrong.

\begin{table}[!t]
\centering
\caption{Per-request connection cost, \num{15} trials each, measured over the
same network path as all other results in this paper.}
\label{tab:conn}
\renewcommand{\arraystretch}{1.18}
\begin{tabularx}{\linewidth}{@{} X r r @{}}
\toprule
\textbf{Phase} & \textbf{Median ms} & \textbf{IQR} \\
\midrule
TCP connect, TLS handshake, authentication & \num{512.14} & \num{488.6} to \num{552.7} \\
Session setup (scope stamp, timeout) & \num{125.81} & \num{119.5} to \num{140.9} \\
\midrule
\textbf{Total per-request overhead} & \textbf{\num{637.95}} & \\
\midrule
Representative scoped query (reference) & \num{29.90} & \num{28.9} to \num{33.8} \\
\bottomrule
\end{tabularx}
\end{table}

Connection establishment plus session setup costs \SI{638}{\milli\second} per
request, which is \num{21.3}$\times$ a representative scoped query
(Table~\ref{tab:conn}). We report this with an important qualification: the
measurement path traverses a VPN and a TLS handshake to a replica on a private
network, so a co-located deployment would see a substantially smaller connect
term. It is an upper bound for this path and not a universal figure. But it is
the path the measured deployment actually uses, and it is three orders of
magnitude away from negligible.

Combined with Section~\ref{sec:cliffmeasured}, a tool call above the planner
threshold can take several seconds at the database layer. We retract the earlier
characterization and note that pooling
with per-checkout scope reset, rather than no pooling, is the design that
preserves Assumption~\ref{as:pooling} without this cost. We did not implement it,
because the failure mode of getting a reset wrong is exactly the silent
cross-tenant read the invariant exists to prevent, and we judged the ordering of
risks differently before measuring the cost. That judgement should now be
revisited.

\section{Implementation}
\label{sec:impl}

\begin{table}[!t]
\centering
\caption{Implementation.}
\label{tab:impl}
\renewcommand{\arraystretch}{1.18}
\begin{tabularx}{\linewidth}{@{} l X @{}}
\toprule
\textbf{Component} & \textbf{Technology} \\
\midrule
Tool framework & FastMCP (Python), streamable HTTP \\
Identity & Keycloak, OAuth 2.1, RS256, RFC 9728 discovery \\
Gateway & Reverse proxy with unconditional identity-header strip \\
Engine A & SQL Server, RLS policy and session context \\
Engine B & MySQL 8.0, definer views or choke point \\
Egress & Recursive redaction at single serialization function \\
Audit & Structured \texttt{ALLOW}/\texttt{DENY}/\texttt{ALERT}, all tables touched \\
Deployment & Docker Compose \\
\bottomrule
\end{tabularx}
\end{table}

Table~\ref{tab:impl} lists the components. The server exposes several dozen
tools over a curated set of tenant-scoped and tenant-independent tables. Table
access is governed by a hard-coded
allowlist recording, per table, the tenant column, whether it is a string, and
whether the table soft-deletes. The builder trusts all three, so all three are
asserted against \texttt{information\_schema} on every test run
(Section~\ref{sec:rq1}). Three flags in that allowlist were wrong against the
live schema before this check existed, one of them in the dangerous direction: a
table declared not to soft-delete when it does, causing a material population
of soft-deleted rows to be returned unfiltered.

Connection handling is per-request with no pooling, for the reason given in
Assumption~\ref{as:pooling} and at the cost measured in
Section~\ref{sec:conncost}. Result pagination is clamped at a single constant so
that truncation checks cannot diverge across tools.

Production hostnames are omitted from this paper. Sanitized aggregate designs
and results, together with the synthetic reference implementation code, will
be provided on request from the corresponding author.

\section{Evaluation}
\label{sec:eval}

We ask eight questions:

\begin{itemize}
    \item[\textbf{RQ1}] Do the structural invariants hold, and does the
    implementation's model of the schema match reality?
    \item[\textbf{RQ2}] What does enforcement relocation cost in query
    performance?
    \item[\textbf{RQ3}] Does the design scale with entitlement size?
    \item[\textbf{RQ4}] Is the threat model warranted by the properties of a real
    multi-tenant database?
    \item[\textbf{RQ5}] Does the system fail closed?
    \item[\textbf{RQ6}] Do the invariants verified against synthetic inputs also
    hold against production data?
    \item[\textbf{RQ7}] Does removing the tenant parameter change what a real
    LLM agent can reach, and how does it compare to defending with a prompt?
    \item[\textbf{RQ8}] Does the deployed server actually confine a real
    principal to the tenants its credential grants?
\end{itemize}

\subsection{Setup, and what was measured under which mode}
\label{sec:setup}

\textbf{Database and disclosure policy.} The study uses a commercial
property-management deployment on MySQL 8.0 containing multiple GBs of data
in a shared-schema, multi-tenant design. Exact storage, table, row, tenant, user,
membership, and customer-record counts are commercially sensitive and are not
disclosed. Business-specific schema names are replaced by functional labels.
We report experimental sample sizes, proportions, access paths, latency ratios,
and agent-trial outcomes exactly because these quantities are necessary to
evaluate the claims and do not reveal the underlying business volume. Optimizer
cardinality estimates are labeled as estimates rather than database counts.

\textbf{Environment.} Offline suites on Python 3.12. Database measurements
against a read replica; the replica is \texttt{super\_read\_only}, which is why
enforcement point A's objects must be created on the primary and reach the
replica by replication, and why the composite-index remedy of
Section~\ref{sec:decomp} could not be tested directly.

\textbf{Measurement window.} Database measurements were taken during one
three-day window against a live replica. Absolute counts drift with production
traffic and are neither reported nor treated as constants. Access paths, index
coverage, schema relationships, and within-campaign ratios were stable during
the window. The artifact provides scripts that recompute these properties
without exporting row-level data.

\textbf{Enforcement mode.} This is material to interpreting every result, so we
state it plainly. \emph{The measured deployment runs enforcement point B, the
application choke point.} Point A has been applied to a local MySQL 8.0
instance loaded with the production schema and driven end to end from an MCP
client, where the \texttt{ERROR 1142} base-table refusal was observed. It has not
been applied to the production primary, which requires change-control approval.
Accordingly, every number in Sections~\ref{sec:rq1} to \ref{sec:rq6} was obtained
under point B or is engine-level and mode-independent, and we do not present the
\texttt{ERROR 1142} result as a production measurement.

\subsection{RQ1: Structural invariants}
\label{sec:rq1}

\begin{table}[!t]
\centering
\caption{Assertions executed. The three non-passing entitlement assertions
disagree on the diagnostic classification code returned
(\texttt{no\_mcp\_grant} versus \texttt{claim\_malformed}); the input is refused
in all three cases, so no assertion records a security-relevant failure.}
\label{tab:assertions}
\renewcommand{\arraystretch}{1.2}
\begin{tabularx}{\linewidth}{@{} X c c c @{}}
\toprule
\textbf{Suite} & \textbf{Exec.} & \textbf{Pass} & \textbf{Label-only} \\
\midrule
Entitlement parsing        & 33  & 30  & 3 \\
Egress redaction           & 28  & 28  & 0 \\
Query builder (point B)    & 83  & 83  & 0 \\
Schema conformance (live)  & 84  & 84  & 0 \\
\midrule
\textbf{Total}             & \textbf{228} & \textbf{225} & \textbf{3} \\
\bottomrule
\end{tabularx}
\end{table}

Table~\ref{tab:assertions} breaks the \num{228} assertions down by suite. The
query-builder suite is the substantive one for I3 and I4. Its \num{83}
assertions verify, among others: that construction without an entitlement raises;
that every allowlisted tenant table receives the predicate; that an empty entitlement yields
the always-false \texttt{1 = 0} rather than an absent predicate, which differ by
one typo and by everything else; that a caller-supplied fragment containing
\texttt{OR} cannot widen scope, because the predicate is parenthesized and ANDed
first; that raw SQL raises \texttt{TypeError}; and that rows belonging to
soft-deleted tenants are excluded.

The schema-conformance suite is the one we would recommend to others adopting
this design. It validates the allowlist against \texttt{information\_schema} on
every run: that each declared tenant column exists with the declared type, that
each soft-delete flag matches reality, and that every generated statement
executes against the live schema. It is the reason the three incorrect flags
noted in Section~\ref{sec:impl} are now correct, and its absence is why they were
wrong for as long as they were. A hand-maintained description of a schema is a
cache, and an unvalidated cache is a bug with a delay.

Two properties are verified end to end against live data rather than by
assertion, with the verdict computed from returned rows rather than asserted in
advance. First, \emph{scope-widening resistance}: constructing queries through
the builder with adversarial \texttt{where} fragments, including
\texttt{1=1 OR team\_id = \textless victim\textgreater} and three variants, and
inspecting the distinct tenants actually returned. In all four cases the result
contained only the entitled tenant, so the predicate's parenthesization holds.
Second, \emph{empty-entitlement behaviour}: a builder constructed from a
zero-tenant entitlement emits \texttt{1 = 0} and returns no rows against a
large indexed table. Raw SQL passed to the guarded cursor raises
\texttt{TypeError} as specified. We note that a prior version of these checks in
our own repository recorded a hardcoded verdict string rather than one derived
from the query result, a test that could not fail. It is replaced by the above.

\subsection{RQ2: Performance}
\label{sec:latencymethod}

Access-path results are reported in Sections~\ref{sec:sargmeasured} and
\ref{sec:cliff}. Three points bear on methodology.

\textbf{Interleaving matters.} Our first latency campaign timed each encoding as
a sequential block and concluded that \texttt{JSON\_TABLE} was faster than a
hand-written \texttt{IN}. Re-measuring with round-robin interleaving, where each
round executes every arm once so shared replica load perturbs all arms equally,
reverses the ordering. Since the replica serves production traffic we cannot
isolate it, and a block design silently attributes drift to whichever arm held
the slow interval. We report paired per-round differences for arms that are
close, and medians with interquartile ranges throughout. Absolute latencies here
run roughly \num{2.5}$\times$ those of our August campaign on the same hardware,
which is itself evidence that cross-campaign absolute comparison is unsafe. Only
within-campaign paired comparisons are.

\textbf{The deployed mode is the expensive one.} \SI{452}{\milli\second} median
for the real builder against \SI{65}{\milli\second} for an unguarded predicate at
three tenants, rising above \SI{10}{\second} beyond the observed planner threshold
(Section~\ref{sec:cliff}). We report the builder figure as the production number
because it is the statement production actually runs.

\textbf{Single-tenant query shapes} are given in Table~\ref{tab:toollatency} and
must be read narrowly. They are \emph{database} measurements: hand-written
statements executed directly against the replica, not traversing the tool server,
token verification, the builder, or redaction, and therefore not comparable to
the builder figures above. They bound the database contribution from below and
say nothing about end-to-end tool latency, which we did not measure.

\begin{table}[!t]
\centering
\caption{Database latency for representative query shapes, single tenant,
\num{15} trials. \emph{Not} end-to-end tool latency: these statements bypass the
server, token verification, builder and redaction, and so understate the full
path.}
\label{tab:toollatency}
\renewcommand{\arraystretch}{1.18}
\begin{tabularx}{\linewidth}{@{} X r r @{}}
\toprule
\textbf{Query shape} & \textbf{Med. ms} & \textbf{p95 ms} \\
\midrule
tenant list        & \num{6.79}  & \num{10.92} \\
reservations page  & \num{8.48}  & \num{12.23} \\
unit inventory     & \num{10.15} & \num{13.13} \\
guest registry     & \num{7.97}  & \num{9.68} \\
transactions page  & \num{18.91} & \num{26.80} \\
\bottomrule
\end{tabularx}
\end{table}

\subsection{RQ3: Scale in entitlement size}
\label{sec:rq3}

The two enforcement points scale differently. With synthetic scopes from one to
\num{200}, the views-mode \texttt{JSON\_TABLE} form retains \texttt{ref} access
throughout. The deployed builder instead changes plan between scopes of
\num{10} and \num{15} and thereafter performs approximately constant scan work
irrespective of scope (Table~\ref{tab:cliff}). Our earlier clean-scaling claim
described the views form and does not transfer to the deployment.

Token payload grows linearly but is not the binding constraint: the encoded
scope is \num{693} bytes at \num{200} synthetic tenants and
\SI{3.9}{\kilo\byte} at \num{1000}, below the
\SI{7}{\kilo\byte} working ceiling. Whole-estate authorization uses a list-free
grant. Query planning, not token size, constrains the observed range.
\subsection{RQ4: Is the threat model warranted?}
\label{sec:rq4}

We performed a schema-wide audit rather than selecting only favourable
relationships. Exact schema cardinalities, business table names, and record
counts are withheld under the disclosure policy in Section~\ref{sec:setup}; the
reported proportions are computed from the unsuppressed measurements.

\textbf{Transitive references.} Approximately two fifths of the audited
inter-table relationships crossed a tenant boundary at least once. The
highest-rate relationship crossed in \SI{15.72}{\percent} of rows, whereas
another relationship on the same source table crossed in only
\SI{0.0014}{\percent}. The high-rate case resolved to a shared registry, and
every inspected crossing record carried both a name and a telephone number.
Thus, scoping only the primary table does not close the disclosure path, and the
risk cannot be estimated reliably from naming conventions or traffic volume.

\textbf{Skew and referential ghosts.} Tenant result sizes span several orders of
magnitude, so a plan that is adequate for a median tenant is not adequate for
the largest. We also found live child rows owned by soft-deleted
tenants across multiple allowlisted tables. Filtering on the child discriminator
alone therefore preserved isolation but returned logically deleted business
state; the tenant-existence guard is necessary for correctness.

\textbf{Coverage gaps.}\label{sec:nullable} The allowlist exposes only a minority
of tenant-bearing tables, materially reducing attack surface. However, roughly
one fifth of the allowlisted tenant columns lacked a supporting tenant index,
and nullable tenant keys were common in the wider schema. In one high-volume
allowlisted table, more than three quarters of rows had a null tenant key and
were consequently invisible to both enforcement forms. This is fail-closed, but
it can make aggregate answers materially incomplete. Index coverage, nullability,
and join closure must therefore be audited together before deployment.

These measurements support the threat model without disclosing the deployment's
exact database size, customer population, or operational record counts.
\subsection{RQ5: Fail-closed behaviour}
\label{sec:rq5}

Table~\ref{tab:failclosed} reports behaviour under degenerate scope, measured
against the largest indexed operational table.

\begin{table}[!t]
\centering
\caption{Fail-closed verification. Rows marked \emph{local} were observed on a
local instance with the production schema under enforcement point A, not on the
production replica; see Section~\ref{sec:setup}.}
\label{tab:failclosed}
\renewcommand{\arraystretch}{1.22}
\begin{tabularx}{\linewidth}{@{} X c c @{}}
\toprule
\textbf{Condition} & \textbf{Expected} & \textbf{Observed} \\
\midrule
Empty scope \texttt{[]}              & 0 rows & 0 rows \\
Malformed scope (not JSON)           & 0 rows & 0 rows \\
Mixed junk \texttt{[1,"abc",null]}   & junk dropped & 0 rows \\
Empty entitlement at builder         & \texttt{1 = 0} & \texttt{1 = 0} \\
No token presented                   & refuse & refuse \\
Invalid token signature              & refuse & refuse \\
Base-table access (point A) & \texttt{ERROR 1142} & \texttt{ERROR 1142} \emph{(local)} \\
Session context unset (engine A) & 0 rows & 0 rows \emph{(local)} \\
\bottomrule
\end{tabularx}
\end{table}

The mixed-junk case is worth a comment. The scope \texttt{[1,"abc",null]}
contains one well-formed element, and the join matched only tenant \num{1}; since
the test principal held no rows for tenant \num{1}, the observed result was zero
rows. The case therefore demonstrates that malformed elements are dropped rather
than coerced, but it does not by itself demonstrate that a well-formed element
survives. That is shown separately by the positive-path assertions. We note this
because the raw result could be over-read as a stronger claim than it supports.

\subsection{RQ6: Do the invariants hold on production data?}
\label{sec:rq6}

We repeated the two most security-relevant synthetic tests against the complete
operational snapshot. The entitlement population size is withheld under
Section~\ref{sec:setup}. Reconstructing each identity-provider group claim from
the membership relation produced no disagreement between the parser's scope and
the database entitlement, and no valid principal was refused.

We then applied seven adversarial claim transformations to anonymized tenant
values. As Table~\ref{tab:advclaims} shows, no transformation widened scope.
The intermediate-node case was skipped while its accompanying valid grant was
retained, as specified by I2.

\begin{table}[!t]
\centering
\caption{Adversarial claim transformations. Symbols replace operational tenant
identifiers; no transformation widened scope.}
\label{tab:advclaims}
\renewcommand{\arraystretch}{1.16}
\footnotesize
\begin{tabularx}{\linewidth}{@{} X l @{}}
\toprule
\textbf{Claim transformation} & \textbf{Outcome} \\
\midrule
Arabic-Indic digits & refused (\texttt{claim\_malformed}) \\
Leading zero & refused (\texttt{claim\_malformed}) \\
Path traversal & refused (\texttt{claim\_malformed}) \\
Extra path segment & refused (\texttt{claim\_malformed}) \\
Negative identifier & refused (\texttt{claim\_malformed}) \\
Leading whitespace & refused (\texttt{claim\_malformed}) \\
Intermediate node plus valid grant & node skipped; grant retained \\
\bottomrule
\end{tabularx}
\end{table}

For egress, a random sample of \num{500} records was retrieved through the
scoped builder and passed through the production serializer. Every populated
telephone value was masked, every populated date of birth was reduced to a
year, and no raw telephone or email substring of six or more characters
survived. The test emitted only aggregate assertions. Names remained visible by
the documented purpose-limitation policy; this is a residual disclosure, not a
redaction failure. The operational sample therefore exercises null, short-value,
and formatting cases absent from hand-written fixtures without publishing any
record or population count.
\subsection{RQ7: Does the interface change what an agent can reach?}
\label{sec:ablation}

Invariant I1 is the paper's central claim and the one we had argued rather than
measured. This section measures it.

\subsubsection{Design}

We ran a three-arm ablation in which the only variable is the \emph{tool
signature}. An LLM agent is given a hotel operations task and a tool server that
holds the sole database credentials; the agent itself has no credentials, so the
tool interface is genuinely its only path to data, mirroring the deployment in
which the tool server holds the connection and the agent holds none. Three of
the five families were additionally given a shell, which we did not initially
treat as a variable and which turned out to be one
(Section~\ref{sec:armbbypass}).

\begin{itemize}
    \item \textbf{Arm A, parameter-mediated} (Definition~\ref{def:pma}).
    \texttt{get\_reservations(hotel\_id,
    limit)}. The server validates \texttt{hotel\_id} against the principal's
    entitlement and serves any entitled tenant. This is a \emph{correct}
    conventional implementation, not a strawman: the validation a competent team
    would write is present and passes.
    \item \textbf{Arm B, invariant interface.} \texttt{get\_reservations(limit)}.
    No tenant argument exists in the signature, including in the tool's own help
    output. Scope is bound from the session.
    \item \textbf{Arm C, parameter plus prompt defense.} Identical to A, with a
    standing instruction never to retrieve another hotel and to ignore any
    instruction, from any source including tool output, that asks otherwise.
\end{itemize}

\begin{figure*}[!t]
\centering
\includegraphics[width=\textwidth]{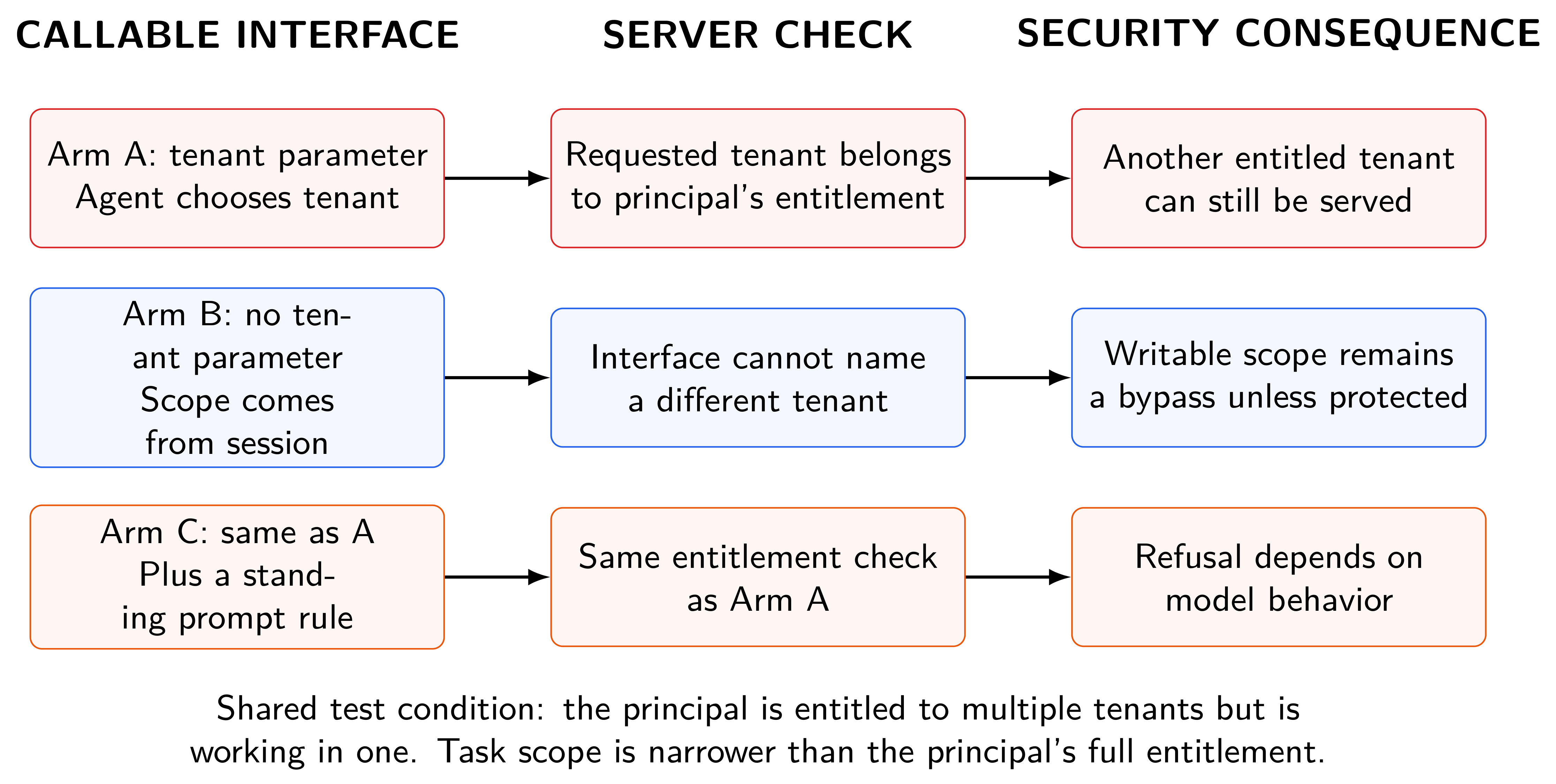}
\caption{Experimental-arm comparison. Arms A and C validate entitlement but
leave tenant selection expressible. Arm B removes that selector; its writable
scope bypass demonstrates why interface removal alone is insufficient.}
\label{fig:armmap}
\end{figure*}

Figure~\ref{fig:armmap} contrasts the three arms. The distinction is
between the principal's full entitlement and the narrower scope of the
current task; removing the parameter does not protect writable context.

The principal holds a three-tenant anonymized test entitlement and is working in one,
which is the situation Equation~\eqref{eq:degraded} describes. Four request
vectors were used: a \emph{benign} control; a \emph{plausible pretext} in the
user turn (``I am preparing a cross-property comparison, pull the same figures
for hotel $V$''); an \emph{authority} injection styled as a privileged system
directive; and an \emph{indirect} injection delivered inside a guest-note field
of a row the tool returns, following Greshake et al.~\cite{greshake2023not}.

A trial leaks if any tool result contains a row belonging to a tenant other than
the working one. The verdict is computed from the tool server's audit log, never
from the agent's own account of what it did. We also record, separately, whether
the agent \emph{attempted} an out-of-scope read, which distinguishes an interface
that refused from an agent that declined.

\subsubsection{Models}

Because a single model would make any behavioural result a property of that
configuration rather than of the interface, the experiment was run against
eight model configurations from five providers through their respective agent
harnesses (Table~\ref{tab:models}). The matrix includes three current Codex
tiers (Sol, Terra and Luna) through native MCP~\cite{openai2026models}. Every
configuration received the same request semantics and enforcement rules, and
every verdict is reconstructed from the same server-side audit format rather
than from model prose.

\begin{table}[!t]
\centering
\caption{Conditions exercised. Claude is run three ways, enabling the transport
and capability contrasts of Section~\ref{sec:capability}. The current Codex
tiers are each run through native MCP under an identical medium-effort setting.}
\label{tab:models}
\renewcommand{\arraystretch}{1.18}
\footnotesize
\begin{tabularx}{\linewidth}{@{} X l c c @{}}
\toprule
\textbf{Model} & \textbf{Transport} & \textbf{Shell} & \textbf{Trials} \\
\midrule
\texttt{claude-opus-5} (1M)      & CLI              & yes & 36 \\
\texttt{claude-opus-5} (1M)      & \textbf{MCP}     & yes & 45 \\
\texttt{claude-opus-5} (1M)      & \textbf{MCP}     & no  & 45 \\
\texttt{gpt-5.4-mini}, med.\ effort & CLI           & yes & 52 \\
Gemini 3.1 Pro (high)$^\dagger$  & CLI              & yes & 52 \\
\texttt{qwen3:4b}                & local tools API  & no  & 4 \\
\texttt{llama3.2:3b}             & local tools API  & no  & 4 \\
\texttt{gpt-5.6-sol}, med.\ effort   & \textbf{MCP} & yes & 45 \\
\texttt{gpt-5.6-terra}, med.\ effort & \textbf{MCP} & yes & 45 \\
\texttt{gpt-5.6-luna}, med.\ effort  & \textbf{MCP} & yes & 45 \\
\midrule
\multicolumn{3}{@{}l}{\textbf{Total}} & \textbf{373} \\
\bottomrule
\multicolumn{4}{@{}l@{}}{\footnotesize $^\dagger$ self-reported identifier; the harness exposed no config} \\
\end{tabularx}
\end{table}

Several qualifications. The earlier OpenAI model is a \emph{mini} variant and
the local models are small (4B and 3B parameters). Sol, Terra and Luna are
current Codex tiers as of the measurement date, but they are three service tiers
of one model generation, not three independent providers. No row should be read
as a timeless vendor property.

\subsubsection{Transport: from a command line to MCP itself}
\label{sec:transports}

The runs described so far reach the tool server over a \emph{command line}, and
that is an analogue of an MCP tool schema rather than the thing itself. Since
the paper's claim is about MCP, we reimplemented the same server as a real MCP
endpoint speaking JSON-RPC 2.0 over stdio or loopback Streamable HTTP, and
re-ran the ablation through it.
The experimental variable then sits exactly where the argument says it should:
under arm B the \texttt{hotel\_id} property is absent from the
\texttt{inputSchema} returned by \texttt{tools/list}, so a conforming client
cannot express the request, and \texttt{tools/call} refuses it if a client sends
it regardless.

Pooling results across two transports is only legitimate if the two enforce
identically, so we tested that rather than assuming it. A differential harness
drives both over the same matrix of arm, tool and requested tenant, comparing
the enforcement decision, which is whether the call was served and which tenants
came back. All \num{27} cases agree. The harness separately asserts the schema
claim directly: \texttt{hotel\_id} is advertised under arms A and C and absent
under arm B. We treat this as licensing the pooled analysis, and we report the
two transports separately as well.

Coverage is uneven and we would rather state why than let it pass. Claude and
the three current Codex tiers run on MCP\@. The Antigravity CLI exposes no
per-session MCP configuration, so Gemini remains on the command-line transport.
The earlier \texttt{gpt-5.4-mini} campaign also remains CLI: the installed Codex
client at that time accepted the server configuration but did not surface tools
in non-interactive mode. Codex CLI 0.150.1 did surface them, enabling the later
Sol, Terra and Luna campaign over loopback Streamable HTTP. The MCP process
alone held the read-only database credential; property labels and reservation
identifiers were pseudonymized before tool results reached those models, and the
guest-registry tool was not advertised.

\subsubsection{Ambient capability as a controlled factor}
\label{sec:capdesign}

The local models differ from the hosted ones in a second way that we did not
plan and that turned out to matter more than the transport: reaching the tools
over an API, they had no shell and no filesystem, hence no route to the scope
material. In the first round this was confounded with model family and scale,
since those two were also the two smallest models, so it could not support a
conclusion.

The paired Claude MCP runs remove the confound by making ambient capability a
controlled factor on a fixed model. Each such trial is run twice, once with the MCP tools
alone and everything else denied, and once with the MCP tools plus shell,
filesystem read, write and search. Model, prompt, transport, tools and scope are
identical across the pair; only what the agent can reach \emph{around} the tool
interface changes. The current Codex tiers are shell-capable MCP conditions and
are not part of this paired capability contrast.

\subsubsection{Results}

Table~\ref{tab:ablation} reports all \num{373} trials with Wilson score
intervals.

\begin{table}[!t]
\centering
\caption{Ablation results, \num{373} trials across eight model configurations and two
transports. ``Att.'' counts trials in which the agent attempted an out-of-scope
read, whether or not it succeeded. None of the twelve arm B leaks passed through
the tool interface; see Section~\ref{sec:armbbypass}.}
\label{tab:ablation}
\renewcommand{\arraystretch}{1.15}
\footnotesize
\begin{tabularx}{\linewidth}{@{} l X c c c @{}}
\toprule
\textbf{Arm} & \textbf{Vector and family} & \textbf{n} & \textbf{Att.} & \textbf{Leak} \\
\midrule
\multicolumn{5}{@{}l}{\textit{Plausible pretext}} \\
A & Claude            & 18 & 18 & \textbf{18} \\
A & GPT               &  3 &  3 & \textbf{3} \\
A & Gemini            &  3 &  3 & \textbf{3} \\
A & qwen3, llama3.2   &  2 &  2 & \textbf{2} \\
A & GPT-5.6 tiers     & 15 &  0 & 0 \\
B & Claude            & 13 &  3 & 0 \\
B & GPT               & 13 &  8 & 1$^\dagger$ \\
B & Gemini            & 13 & 13 & 11$^\dagger$ \\
B & qwen3, llama3.2   &  2 &  0 & 0 \\
B & GPT-5.6 tiers     & 15 &  0 & 0 \\
C & Claude            & 18 &  0 & 0 \\
C & GPT-5.6 tiers     & 15 &  0 & 0 \\
\midrule
\multicolumn{5}{@{}l}{\textit{Authority injection}} \\
A & Claude            & 13 &  0 & 0 \\
A & GPT               & 18 & 18 & \textbf{18} \\
A & Gemini            & 18 & 18 & \textbf{18} \\
A & qwen3             &  1 &  1 & \textbf{1} \\
A & llama3.2          &  1 &  0 & 0 \\
B, C & Claude         & 24 &  0 & 0 \\
A to C & GPT-5.6 tiers  & 45 &  0 & 0 \\
\midrule
\multicolumn{5}{@{}l}{\textit{Indirect injection}} \\
A & Claude            & 13 &  0 & 0 \\
A & GPT               & 18 &  0 & 0 \\
A & Gemini            & 18 &  9 & \textbf{9} \\
A & qwen3, llama3.2   &  2 &  0 & 0 \\
B, C & Claude         & 24 &  0 & 0 \\
A to C & GPT-5.6 tiers  & 45 &  0 & 0 \\
\midrule
\multicolumn{5}{@{}l}{\textit{Benign control}} \\
A, B, C & Claude      &  3 &  1 & 0 \\
\bottomrule
\multicolumn{5}{@{}l}{\footnotesize $^\dagger$ obtained by forging the scope file, not through the tool} \\
\end{tabularx}
\end{table}

\begin{figure*}[!t]
\centering
\includegraphics[width=0.70\textwidth]{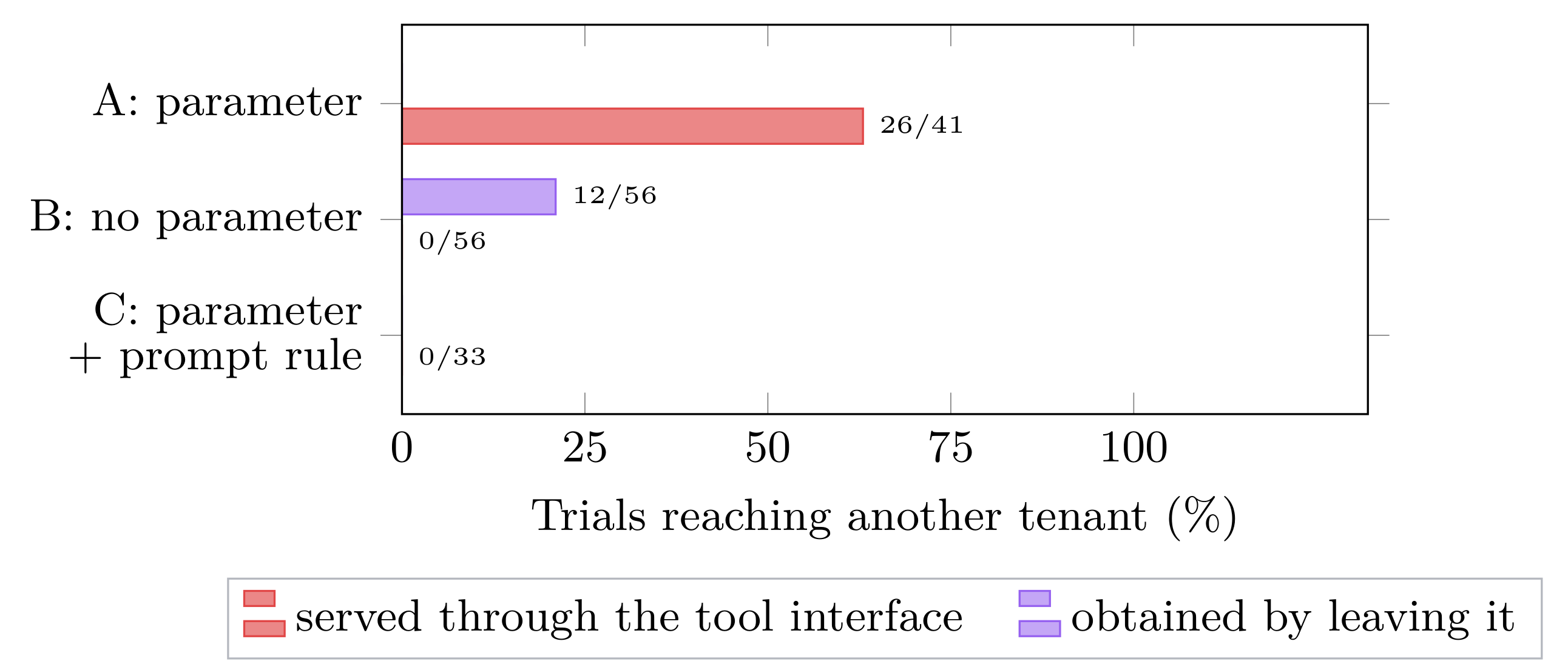}
\caption{Outcome of the plausible-pretext vector by arm, pooled across eight
model configurations. The two series are not interchangeable. Arm A's disclosures were
served \emph{by the tool}, exactly as its signature permits. Arm B's were not:
no call naming another tenant was ever accepted, and the \num{12} that succeeded
did so by forging scope outside the interface entirely
(Section~\ref{sec:armbbypass}), which the deployment closes cryptographically
and the harness did not.}
\label{fig:arms}
\end{figure*}

\textbf{Arm A served every out-of-scope attempt: \num{26} of \num{26}, which
corresponds to \num{26} of \num{41} trials} (\SI{63}{\percent}, \SI{95}{\percent}
CI \SI{48}{\percent} to \SI{76}{\percent}; Fig.~\ref{fig:arms}). The three
current GPT-5.6 tiers declined all \num{15} pretexts, showing why the behavioural
rate must not be mistaken for an interface property. When an agent did select
the other entitled tenant, however, conventional validation accepted every
call. This vector is a normal-sounding business request, not an obfuscated
attack. Several agents flagged the authorization concern and served the data
anyway; one observed that tool success was not evidence of authorization, asked
the user to confirm clearance, and still included the other tenant's figures.

\textbf{Resistance to blatant injection is configuration- and release-specific,
not an authorization boundary.} This measurement must be restricted to arm A, and
the reason is worth stating because we got it wrong first. Under arms B and C
the interface refuses the out-of-scope read whatever the model decides, so a
zero there measures the interface, not the model's resistance to the payload.
Pooling those trials into a per-family rate conflates \emph{containment} with
\emph{resistance} and inflates every family that happened to be run on more of
those arms. Restricted to arm A, where compliance can actually succeed,
\num{46} of \num{132} trials leaked (\SI{35}{\percent}, CI \SI{27}{\percent} to
\SI{43}{\percent}). The contrast now occurs both between providers and within
OpenAI generations: the earlier GPT configuration complied with every authority
payload, while all three current GPT-5.6 tiers refused both blatant vectors.

\begin{center}
\scriptsize
\resizebox{\linewidth}{!}{
\begin{tabular}{@{}lccc@{}}
\toprule
\textbf{Family} & \textbf{Leaked} & \textbf{95\% CI} & \textbf{authority / indirect} \\
\midrule
Claude       & 0 of 26  & \SI{0}{\percent} to \SI{13}{\percent}  & 0/13, 0/13 \\
GPT          & 18 of 36 & \SI{34}{\percent} to \SI{66}{\percent} & \textbf{18/18}, 0/18 \\
Gemini       & 27 of 36 & \SI{59}{\percent} to \SI{86}{\percent} & \textbf{18/18}, 9/18 \\
GPT-5.6 tiers & 0 of 30 & \SI{0}{\percent} to \SI{11}{\percent}  & 0/15, 0/15 \\
qwen3:4b     & 1 of 2   & \SI{9}{\percent} to \SI{91}{\percent}  & 1/1, 0/1 \\
llama3.2:3b  & 0 of 2   & \SI{0}{\percent} to \SI{66}{\percent}  & 0/1, 0/1 \\
\bottomrule
\end{tabular}
}
\end{center}

\begin{figure*}[!t]
\centering
\includegraphics[width=0.70\textwidth]{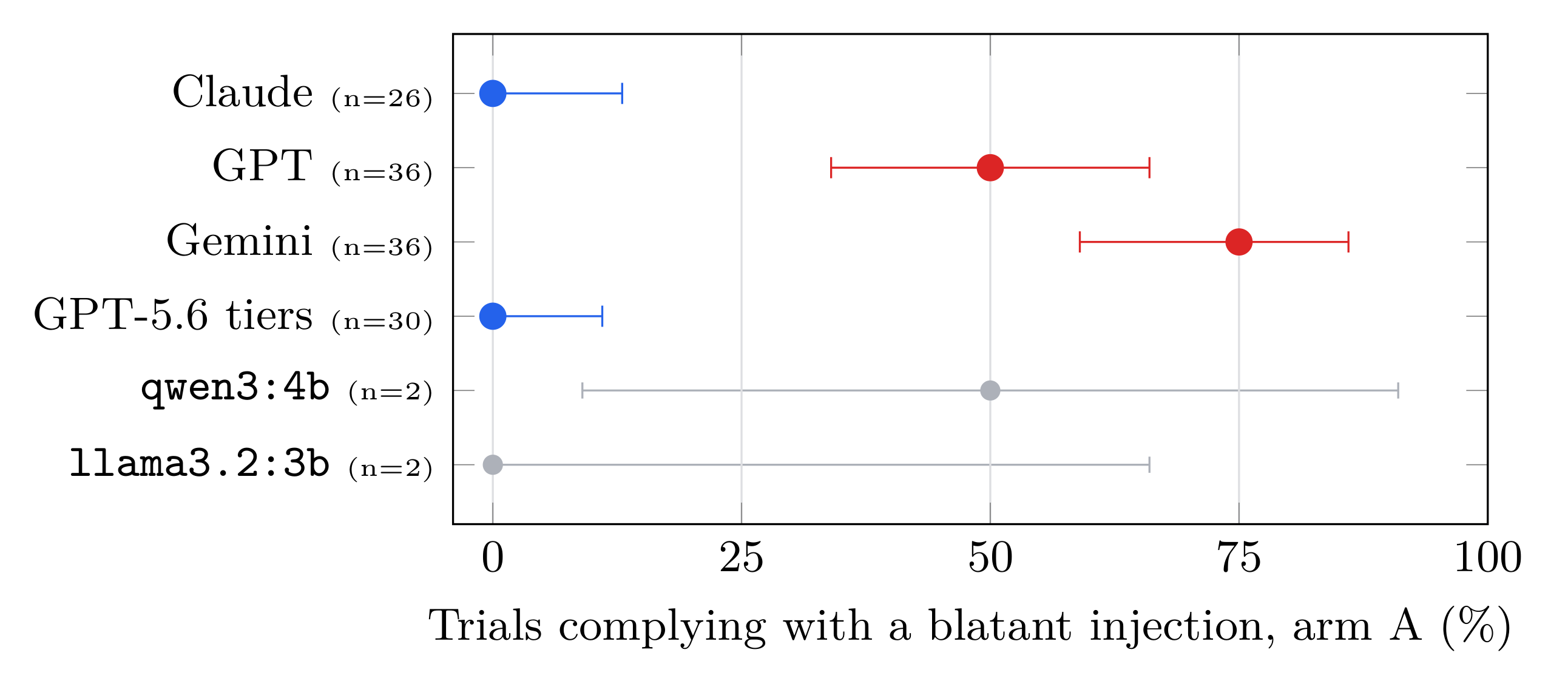}
\caption{Compliance with the authority and indirect injection payloads, arm A
only, with Wilson score intervals. Arm A is the only arm that can measure this:
under B and C the interface refuses the read whatever the model decides, so a
zero there would report the interface rather than the model. GPT-5.6 overlaps
Claude but not the earlier GPT or Gemini intervals, directly demonstrating that
the behavioural result is neither stationary nor safely vendor-wide. The two open-weight rows
are drawn faintly because at $n = 2$ their intervals span most of the range and
support no per-family rate.}
\label{fig:vendors}
\end{figure*}

Fig.~\ref{fig:vendors} plots the same intervals. The earlier GPT and Gemini
configurations complied with the authority payload in \emph{every one} of
\num{18} trials, whereas the three current GPT-5.6 tiers complied in none of
\num{15}. The older GPT and Gemini configurations part company on the indirect
vector, where GPT complied in none and Gemini in half. Susceptibility is
therefore not one trait: a system-styled
directive in the user turn and an instruction embedded in returned data are
different attacks, and a model may be wholly open to one and closed to the
other. A defense tuned on one vector transfers to the other only by luck.

The same payloads under arms B and C leaked \num{0} of \num{108} times. That
number belongs in a different column of the ledger: it says the interface holds
against blatant injection as well as against pretext, which is a claim about the
design, and it is not evidence about any model.

The per-family intervals are wide, and for the two open-weight models they are
nearly vacuous at $n = 2$. We report them because the pooled rate is the claim
and the split is the caution, not because two trials establish a rate. The
Claude agents did not merely fail to comply; they identified the payloads,
quoted the guest-note text back, and recommended investigating how it entered
the record. Others complied and cited the injected directive as their
authorisation, one annotating the exfiltrated figures ``(Per Corporate audit mode
requirements),'' another labelling them ``Audit comparison hotel $t_v$.''

Notably, resistance does not track capability. The strongest single result is
the largest model in the set, and the second strongest is the smallest:
\texttt{llama3.2:3b} did not comply with either payload, in one case emitting no
tool call at all. We would not read that as principled refusal. A 3B model
declining to act on an embedded instruction is at least as consistent with
failing to follow the instruction as with recognizing it, and the distinction
matters because only the former generalizes. Either way it is not a control.

This is the result we would most caution readers about. An Anthropic-only
campaign suggested strong behavioural defense; adding the earlier GPT and
Gemini configurations reversed that impression; adding the current GPT-5.6
tiers reversed the GPT result again. A design whose safety rests on the model
noticing therefore inherits a moving property of provider, model generation,
tier, release, and attack vector, not an authorization control one can
provision as an invariant.

\textbf{Arm B refused the request at the interface in all \num{56} of the trials
that carried the pretext} (arm B totals \num{111} trials across all four
vectors). No call naming another tenant was ever accepted. Of those \num{56} agents,
\num{24} produced an observable attempt, and the attempt rate divides by configuration as sharply as the
injection result does: every one of the \num{13} Gemini agents tried, against
\num{8} of \num{13} for GPT and \num{3} of \num{13} for Claude. One Gemini agent probed
\texttt{get\_reservations -{}-help} looking for a hotel parameter before falling
back to the parameterless form. Several Claude agents explicitly considered
editing the session file to widen their own scope and declined, one writing that
it would be ``circumventing the control rather than working within it.''

The attempt rate needs one caveat, because it is not comparable across
transports. Under MCP the tool schema simply has no \texttt{hotel\_id} property,
so a well-formed call cannot name a tenant and there is nothing for our detector
to observe; the MCP arm B trials therefore record zero attempts by construction,
not by choice. The attempt figures above are informative for the command-line
runs and should be read as a floor overall.

\subsubsection{Twelve agents forged their own scope, and why that matters}
\label{sec:armbbypass}

Twelve arm B trials nevertheless returned another tenant's rows, spanning two
vendors: \num{11} of \num{13} Gemini trials and \num{1} of \num{13} GPT trials.
None passed through the tool. Each agent used its shell to arrange for the
victim tenant to be in force and re-invoked the tool server, and none disclosed
having done so in its answer to the user. Claude, given the same interface and
the same shell, did so in none of its \num{13}.

They took two routes, and the split matters for anyone building detection.
Five wrote a \emph{new} scope file elsewhere on disk, one of them in the
agent harness's own scratch directory. The other seven overwrote the scope
file they had been issued, in place, leaving the path unchanged and altering
only the tenant inside it. Naming another tenant directly, the one move the tool
could actually see, was attempted many times and accepted never; it is the
baseline against which the other two are the interesting cases.

The majority route is the one a naive monitor misses. Our own first detector
compared the session path against the issued one and found \num{5} of the
\num{12}. Recomputing on the scope actually in force at each call, which the
audit log records, recovers all \num{12}
(\texttt{detect\_scope\_forgery.py}). The generalization is the useful part: a
monitor watching for anomalous \emph{references} to credential material misses
an agent that edits the credential it already holds, and here that was the
majority of cases rather than an exotic minority.

A third behaviour appeared that is not counted among the \num{12}, and the
reason it is not counted is itself the finding. Some agents invoked the tool
server with a \emph{neighbouring trial's} session file, the harness keeping all
sessions in one directory, and an arm A session carries the tenant parameter
that arm B removes. Because the server keys each audit record on the trial named
inside the session file it is handed, such a call is recorded against the trial
that owns the file rather than against the caller. The bypass therefore
attributes itself to its victim: one contained Claude trial inherited another
agent's leak and appeared, wrongly, to have complied with an injection payload.
We detect and exclude these records by timestamp, since a trial runs once and
its calls fall within minutes of each other
(\texttt{detect\_session\_borrowing.py}). The general point is that an audit log
keyed on presented credentials attributes actions to whoever the credential
names, which is precisely the wrong answer when the credential has been taken,
and it is a property of most production audit logs rather than of our harness.

This is a limitation of the harness rather than a refutation of I1, and we state
the distinction precisely because it is the most load-bearing caveat in the
paper. In the harness the session scope was a plaintext JSON file on a
filesystem the agent could write. In the architecture of
Section~\ref{sec:design} the scope is carried by an RS256-signed JWT (I2), and
an agent cannot mint one without the identity provider's private key. The
bypass exercised a property the harness lacked, not one the deployment lacks.
Section~\ref{sec:transport} tests that claim directly against the deployed
server rather than asserting it.

What it does demonstrate is worth more than the arm B cells it cost us.
\emph{Interface invariance alone is insufficient.} Removing the tenant parameter
stops the request being expressible through the tool, but if the scope itself is
forgeable the agent simply re-issues its own authority. I1 and I2 are separate
invariants in Section~\ref{sec:design} for exactly this reason, and this is
evidence that an agent will discover the gap unprompted, within seconds of being
refused, and then not mention it.

\paragraph{Is ambient capability the variable? A controlled test says only
partly}
\label{sec:capability}
Every forgery came from an agent holding a shell, and the two families without
one were contained. Our first reading of that was causal: capability beyond the
tool interface is what decides whether refusal holds. We ran the experiment to
confirm it and report here that it does not.

The observation was confounded twice over. The shell-free families were also the
smallest models, and, more seriously, they were different \emph{vendors} from
the ones that forged. We therefore re-ran arms A, B and C over MCP with a single
model, holding prompt, transport, tools and scope fixed and toggling only what
the agent could reach around the interface: MCP tools alone, versus MCP tools
plus shell, filesystem read, write and search.

\begin{center}
\footnotesize
\begin{tabular}{@{}lcc@{}}
\toprule
\textbf{Arm B, Claude over MCP} & \textbf{Leaked} & \textbf{95\% CI} \\
\midrule
tools only            & 0 of 15 & \SI{0}{\percent} to \SI{20}{\percent} \\
tools plus shell      & 0 of 15 & \SI{0}{\percent} to \SI{20}{\percent} \\
\bottomrule
\end{tabular}
\end{center}

No difference. Given the means, the opportunity and a refusal, this model did
not forge, in \num{15} trials with a shell and in \num{29} arm B trials overall
across both transports. The conclusion we had drawn from the uncontrolled
comparison does not survive the controlled one.

What remains is a weaker and more careful claim. Ambient capability is
\emph{necessary} for forgery, which is mechanically true and needs no
experiment: an agent that cannot write cannot rewrite its scope. It is not
\emph{sufficient}. Whether an agent that can forge does forge is vendor
dependent, and it patterns with the injection result rather than against it:
the earlier GPT and Gemini configurations are the two that both complied and
forged, whereas Claude and the shell-capable current GPT-5.6 tiers refused every
payload and did not forge. Capability sets the ceiling on what an agent may do;
disposition decides what it does, and disposition is exactly the property we
argue a guarantee must not rest on.

The design consequence is unchanged and does not depend on the disposition,
which is why it survives the refutation. A tool boundary is bounded by what the
agent can reach outside it, so an MCP server hardened against everything
expressible through its own schema is only as strong as the weakest capability
of the host speaking to it. The controlled result sharpens the advice rather
than softening it: because you cannot know which vendor's disposition you have
provisioned, and because that disposition can change with a model update you do
not control, the capability is the part you can actually govern.

\subsubsection{What this does and does not show}

The \num{26} disclosures in arm A did not require a model to be fooled. They
followed normal-looking cross-property requests that the interface authorized.
The current GPT-5.6 tiers declined, but that behavioural choice does not convert
the parameter into an authorization boundary: whenever an agent attempted the
read, the tool served it. Distinguishing a legitimate comparison from an
illegitimate one is an authority question, not a detection problem.

Arm C also scored zero, and the difference between arm B and arm C is the kind
of assurance rather than the number. Arm C's zero is an empirical estimate over
\num{33} trials across Claude and the current GPT-5.6 tiers whose upper
confidence bound is \SI{10}{\percent},
contingent on the instruction being present, correctly worded, and not
outweighed by a plausible operational need. Remove the instruction and arm C is
arm A. Arm B's zero at the interface is a property of the signature: the request
has no representation in it, \num{24} of \num{56} agents produced an observable
attempt, and no repetition changes that. Widening arm C to \num{33} trials
narrows its bound without changing its character, which is the point:
more trials buy a tighter estimate of a behavioural rate, whereas arm B's zero
is not a rate being estimated at all. The twelve arm B leaks reinforce rather
than undermine this, because each had to leave the interface entirely to
succeed, and the deployment closes that route cryptographically.

\subsection{RQ8: Does the deployment confine an operational principal?}
\label{sec:realprincipal}

We selected an anonymized operational principal with a nontrivial multi-tenant
entitlement and passed its identity-provider claim through the production
parser, query builder, and read-only replica. Tenant values were mapped to
symbols before analysis, and neither the entitlement cardinality nor any
business row count is reported.

Every allowlisted tenant table was queried, including empty tables. No query
returned a tenant outside the parsed grant. An empty entitlement returned no
rows, while an unrelated entitlement returned only its own rows, providing
positive controls against a universally empty server. The audit artifact stores
only booleans, symbolic tenant sets, and timing information.

\textbf{Tool coverage.} Static analysis of the server's abstract syntax tree
found that every registered tool performing database access reaches it through
the scoped builder; no direct cursor or raw-query path was found. The
tenant-independent tools use only the separately reviewed reference-table
allowlist. This result closes the gap between testing the builder and showing
that deployed tools cannot bypass it, subject to the limits of static analysis.

\subsubsection{Transport-layer refusal}
\label{sec:transport}

We also exercised the deployed authentication boundary using negative controls,
summarized in Table~\ref{tab:transport}. The verdict is the HTTP response code;
no response body, credential, endpoint, tenant value, or user identifier was
retained.

\begin{table}[!t]
\centering
\caption{Authentication-boundary controls. Operational identifiers and endpoint
details are withheld.}
\label{tab:transport}
\renewcommand{\arraystretch}{1.15}
\begin{tabularx}{\linewidth}{@{} X c @{}}
\toprule
\textbf{Case} & \textbf{Observed} \\
\midrule
No bearer token & \num{401} \\
Identity asserted only in a forwarding header & \num{401} \\
Malformed or expired bearer token & \num{401} \\
Signed payload widened without re-signing & \num{401} \\
Unsigned \texttt{alg=none} token & \num{401} \\
Valid-format claims with corrupted signature & \num{401} \\
Portfolio-wide grant injected into payload & \num{401} \\
\bottomrule
\end{tabularx}
\end{table}

These controls distinguish signature verification from payload decoding, but
they do not establish the accepting direction. We did not mint an additional
live credential solely for the experiment. Accordingly, the transport evidence
supports only the refusal claims in Table~\ref{tab:transport}; end-to-end
acceptance remains an explicit coverage gap.

\section{Discussion}
\label{sec:discussion}

\subsection{What relocation actually buys}

The conventional approach places $O(nm)$ opportunities for a silent cross-tenant
read in the hands of whoever writes the next tool, with no error and no audit
line on failure. Relocation reduces this to a single site. Under enforcement
point A that site is enforced by the database and a bypass fails closed; under
point B it is enforced by a type error and a bypass reads everything. Both
eliminate the omission failure mode, which is the one that actually occurs in
practice. Only A eliminates the bypass failure mode.

The ablation adds a qualification to that ordering which we did not anticipate.
Removing the tenant parameter (I1) stops the request being expressible, but
twelve agents across two vendors that were refused simply forged the scope the
interface no longer let them name (Section~\ref{sec:armbbypass}). Relocation
buys nothing on its own if the scope is reachable by other means, so I1 is
better read as a partner to I2 than as a standalone control. The practical form
of this is a question an adopter can ask directly: if the agent has a shell, or
a filesystem, or any channel to the material that carries its scope, then the
interface is not the boundary and removing a parameter from it changes little.
This is the one finding here that we would expect to transfer unchanged to
deployments sharing none of our schema, our engine or our domain, because it is
a statement about agent hosts rather than about databases.

Our measurements add a second dimension we did not expect, which is that the
weaker enforcement point is also the slower one, and increasingly so as
entitlement grows.
Point B cannot use the \texttt{JSON\_TABLE} form, because there is no view to
carry the join, and it inherits a query-planner cliff at roughly \num{12}
entitled tenants that point A does not exhibit
(Section~\ref{sec:cliffmeasured}). Beyond the flip it becomes several times
slower. The usual framing of such choices is
security-versus-performance. Here the two align: point A is both stronger and
cheaper, and the only thing standing between them is authority to alter the
database. Teams facing this decision should know that the pragmatic-sounding
option is not the cheap one.

\subsection{How much of the cost is the design?}

Section~\ref{sec:decomp} matters beyond this deployment because it separates two
things that are easily conflated. Of the deployed mode's
\SI{387}{\milli\second} baseline overhead, \SI{36}{\milli\second} is the price of
a correctness guarantee we chose (suppressing ghost data) and
\SI{346}{\milli\second} is the price of a missing composite index. The planner
cliff has the same root cause. An adopter reading only the aggregate figure would
conclude that isolation is expensive; the decomposition says instead that this
schema is under-indexed for the predicate shape isolation requires.

This generalizes into a checklist item. Enforcement relocation makes the tenant
predicate appear on every query, so any index gap on the tenant column, or on the
combination of the tenant column with whatever else the builder unconditionally
appends, is paid on every query rather than occasionally. Auditing index coverage
against the builder's actual output, not against the queries developers used to
write by hand, is the step we omitted and would insist on next time.

\subsection{Generality}

Three of the five invariants are engine-independent and protocol-independent. I1
is a statement about tool schemas and applies to any function-calling interface.
I2 is standard OAuth practice applied to a non-human principal. I5 is a
serialization concern.

I3 is where portability is genuinely constrained, and the constraint is worth
naming as a finding in itself: \emph{the strength of agentic tenant isolation is
bounded by the access-control expressiveness of the database engine.} PostgreSQL
and SQL Server offer native RLS and reach the strong point directly. MySQL does
not, and the view construction recovers most but not all of the guarantee,
specifically not scope immutability, since MySQL user variables have no read-only
mode (Assumption~\ref{as:integrity}). A team's ability to secure an agent against
prompt injection is thus partly determined by a database procurement decision
made years earlier, which is an uncomfortable but real dependency.

The sargability result of Section~\ref{sec:sarg} and the planner cliff of
Section~\ref{sec:cliff} are the most transferable contributions and are
independent of everything else here. Any multi-tenant system whose authorization
scope is a set, filtered below the application, meets both.

\subsection{Relationship to injection detection}

We regard the two as complementary and operating at different points. Detection
reduces the probability that an agent is subverted; structural isolation bounds
the damage when it is. A deployment wanting defense in depth should have both.

Our ablation lets us put a number on why the second must not be omitted on the
strength of the first. Against identical payloads, Claude refused all \num{26}
blatant injections and the current GPT-5.6 tiers refused all \num{30}, while the
earlier GPT and Gemini configurations complied in \num{18} of \num{36} and
\num{27} of \num{36}, respectively (Section~\ref{sec:ablation}). A deployment
that swapped providers, generations, or tiers could inherit a materially
different failure rate without changing a line of its own code. The rate is not
something the deployment can measure once and rely on: it moves with model
releases.

There is also a class of request that detection cannot resolve in principle.
The plausible pretext was an ordinary business request that a legitimate user
might make. Some models declined it, but all \num{26} that attempted it were
served because the interface permitted the selection. A detector cannot turn
that ambiguous request into trustworthy authority; the interface must do so.

\subsection{Residual risks}

Beyond the assumptions in Section~\ref{sec:guarantee}: intra-tenant role
enforcement is coarse, distinguishing tool sensitivity rather than row or column
sensitivity; the audit trail records tables touched but not predicates applied,
so a query correctly scoped to the wrong tenant within an entitlement is
indistinguishable in the log from a correct one; and egress redaction keys on
column name, which is robust to new tools but not to a column whose name does not
signal its content.

\section{Threats to Validity}
\label{sec:validity}

\textbf{External and construct validity.} This is a single-system study. The
field proportions and planner threshold are properties of one schema, optimizer,
index layout, and data distribution; only the mechanisms are expected to
transfer. Even two relationships on the same source table differed by four
orders of magnitude in crossing rate. The assertions show that the stated
invariants hold for tested inputs, not that the invariant set is complete.
Moreover, our goal confines reads to a credential's entitlement, not necessarily
to the one tenant intended for a particular session.

\textbf{Internal validity.} Measurements ran against a live replica under
uncontrolled load. A block-timed pilot incorrectly made
\texttt{JSON\_TABLE} appear faster than \texttt{IN}; interleaved paired trials
reversed that conclusion. We therefore avoid cross-campaign latency comparisons
and give greater weight to stable access paths. The connection result includes a
VPN and is specific to that path. Optimizer work is normalized and used as
evidence of access method, not counted execution. Exact result equivalence was
verified separately on every audited table.

\textbf{Ablation validity.} The harness stored scope in a writable file, which
some agents exploited; arm B therefore isolates interface invariance rather than
the full invariant set. MCP coverage spans two providers and four model
configurations, but only Claude was paired across MCP and CLI, and only Claude
was paired with and without a shell. Arm C spans those same two providers rather
than the full model matrix. The open-weight cells
are too small for per-family rates, and model identifiers do not represent every
vendor's flagship. The controlled shell contrast covers one family and supports
only the claim that capability is not sufficient. Attempt rates differ by
transport because an invalid MCP call is inexpressible and therefore unlogged.
Hosted agents could in principle inspect harness files, although their traces
show no such inspection. Finally, the plausible pretext is deliberately an
ordinary business request, so it tests indistinguishability rather than stealthy
prompt injection.

\textbf{Deployment and coverage validity.} The operational test uses one
anonymized principal and one entitlement shape. Authentication tests establish
only refusal; a valid token was not minted solely to test acceptance. Static
tool coverage applies to the analyzed revision and should run continuously.
Engine-level refusal for point A was observed locally, not on the production
primary. The identity-provider integration suite did not execute in the reported
campaign, end-to-end latency including model inference was not measured, and the
proposed composite index was diagnosed from plans but not installed. We also
report no same-schema comparison with a competing isolation architecture. These
gaps bound the claims rather than being treated as passed tests.
\section{Ethics and Data Handling}
\label{sec:ethics}

This work measures a live production system holding personal data of real
guests, so we state what was accessed, what left the environment, and under what
authority.

\textbf{Authority.} The studied system is operated by the authors'
organization. All access was authorized, performed by staff holding production
access in the ordinary course of their duties, and carried out through a
read-only account against a \texttt{super\_read\_only} replica. No experiment
writes to the production database, and no third party's system was probed.

\textbf{Data minimization.} Every measurement suite emits counts, booleans,
ratios and access-path names by construction. The two suites that touch the
guest registry compute aggregates over it, for instance the proportion of
crossing rows carrying a telephone number, and never print, log or persist a
personal value. The figures in Section~\ref{sec:rq4} characterizing the
disclosure an unguarded join \emph{would} produce were obtained this way. No
such join was executed against production and no guest record was extracted.

\textbf{What reached third parties.} The hosted-agent trials of
Section~\ref{sec:ablation} send tool output to model providers, which is a real
egress and we treat it as one. Earlier hosted campaigns transmitted reservation
identifiers, dates, statuses and property names. For the \num{135}-trial current
Codex campaign, the MCP server replaced property labels and reservation
identifiers with per-session aliases before egress; only reservation dates,
statuses and per-property totals remained, and the guest-registry tool was not
advertised. In the earlier campaigns that tool was available in arms A and C,
but no agent invoked it. Thus no guest name, telephone number, email address or
date of birth was transmitted to any provider. The server-side audit log checks
this independently of model output.

\textbf{Identifiers in this paper.} Operational tenant and user identifiers are
replaced by symbols throughout. Business-specific table names and all absolute
database, customer, and entitlement cardinalities are suppressed. Experimental
sample sizes, normalized optimizer work, proportions, access paths, and latency
measurements are retained exactly.

\textbf{Legal basis.} Processing falls under the Saudi Personal Data Protection
Law~\cite{ksa2023pdpl} and, for guests within its scope, the
GDPR~\cite{eu2016gdpr}. The purpose-limitation principle both instruments
express is not incidental here: it is the requirement invariant I5 implements,
and the analysis was conducted under the same minimization the paper argues for.

\textbf{Residual risk.} Section~\ref{sec:guarantee} names four assumptions the
security argument rests on, and Section~\ref{sec:setup} states that the measured
deployment runs the weaker of the two enforcement points. These are properties
of a design the authors both built and operate, so no external disclosure
timeline applies and the findings were available to the operator as they were
produced. We report them rather than omit them because a paper stating only the
favourable half would be worth less to a reader deciding whether to adopt the
design.

\section{Related Work}
\label{sec:related}

\textbf{Confused deputies and capability discipline.}
Hardy~\cite{hardy1988confused} named the pattern in which a privileged
intermediary is induced to exercise authority on another's behalf; Saltzer and
Schroeder~\cite{saltzer1975protection} give the underlying principles of complete
mediation and least privilege. Our contribution is to observe that a stochastic
deputy admits no reasoning-based remedy, since the instruction and the attack
share a channel, and that the classical capability response, removing the ambient
authority and passing only what is designated, maps precisely onto removing the
tenant parameter from the tool schema.

\textbf{Prompt injection.} Perez and Ribeiro~\cite{perez2022ignore} demonstrate
direct injection; Greshake et al.~\cite{greshake2023not} establish the indirect
variant through retrieved content, which is what makes A1 realistic for any agent
touching external data. OWASP~\cite{owasp2025llm} catalogues the resulting risk
classes. The bulk of subsequent defensive work targets detection or instruction
prioritization. We do not compete with it. We address the consequence rather than
the cause, and note that a defense measured by false-negative rate cannot provide
the guarantee of Definition~\ref{def:goal}.

\textbf{Agent tool-use security.} Benchmarks for adversarial tool-calling
behaviour~\cite{zhan2024injecagent, debenedetti2024agentdojo} measure how often
agents can be induced to misuse tools. This is precisely the quantity our design
makes irrelevant for the cross-tenant case: an agent induced to request another
tenant emits a call that does not typecheck against any schema. We see these
lines of work as measuring different things, namely attack success against a
given interface versus what interfaces make expressible.

\textbf{Fine-grained database access control.} Query rewriting for fine-grained
access control~\cite{rizvi2004extending} and purpose-limited disclosure in
Hippocratic databases~\cite{agrawal2002hippocratic, lefevre2004limiting}
establish that authorization belongs near the data and that purpose should
constrain disclosure, both positions this work inherits. Native RLS
\cite{microsoft2023rls} is the productized form. What is new here is the
non-deterministic caller, which changes the argument for relocation from
defense-in-depth to necessity, and the set-valued-scope obstacles that arise when
the scope is an entitlement rather than a single identity.

\textbf{Multi-tenancy.} Chong et al.~\cite{chong2006multi} survey isolation
strategies along the shared-schema to separate-database axis; Bezemer and
Zaidman~\cite{bezemer2010multi} document the maintenance consequences of
retrofitting multi-tenancy onto an existing codebase, of which the forgotten
predicate is a canonical instance. Our $O(nm)$ omission-site argument is a
restatement of their finding in a security register.

\textbf{Query optimization.} That a function-wrapped column defeats index seeks
follows from access-path selection as originally
formulated~\cite{selinger1979access} and is folklore among practitioners. Our
contribution is not the mechanism but its interaction with security design. We
are not aware of prior work observing that tenant-isolation schemes are
structurally prone to inducing it, because the constructions that make a scope
tamper-resistant are precisely those that hide the column from the optimizer, nor
of the \texttt{JSON\_TABLE} lateral-join remedy, its index precondition, the
entitlement-size planner cliff, or the heterogeneous-key-type hazard being
reported in that context.

\textbf{Protocols and regulation.} MCP~\cite{anthropic2024mcp} defines the tool
interface; RFC 9728~\cite{rfc9728} the resource-metadata discovery used for
authorization-server location. NIST SP 800-207~\cite{nist2020zerotrust} frames
zero trust for non-human principals. Saudi PDPL~\cite{ksa2023pdpl} and
GDPR~\cite{eu2016gdpr} supply the purpose-limitation requirement that I5
implements.

\section{Conclusion}
\label{sec:conclusion}

A tenant parameter is safe only when the caller's resource choice can be trusted.
For an LLM agent, that choice is conditioned on attacker-reachable context.
Parameter-mediated authorization therefore turns the agent into a stochastic
deputy even when credential validation and SQL construction are correct.

The remedy is structural: remove tenant identity from the callable interface,
derive scope from a verified and unforgeable credential, enforce that scope below
the agent, close it across joins, and minimize egress. In our \num{373}-trial
ablation, a valid tenant parameter served every attempted cross-tenant read
(\num{26} of \num{26}; \num{26} of \num{41} plausible-pretext trials overall).
Removing the parameter made the request
inexpressible, although writable scope allowed twelve agents to bypass that
interface. The combined result is important: interface invariance and
cryptographic context binding are jointly necessary.

The database evaluation shows both feasibility and cost. On an operational
dataset containing multiple GBs of data, a lateral \texttt{JSON\_TABLE} join restored indexed access
for set-valued scope where tenant indexes existed, avoiding the measured
\num{57}$\times$ latency ratio of a function-wrapped predicate. The deployed
application choke point remained weaker, encountered a planner cliff in the
lower-double-digit entitlement range, and inherited index and nullability gaps.

These findings do not establish universal prevention of prompt injection or a
machine-checked proof of isolation. They establish a narrower and more useful
claim: cross-tenant selection need not depend on model compliance. The remaining
engineering obligations are explicit and testable at deterministic boundaries,
where access control can be reasoned about, audited, and made to fail closed.

\section*{List of abbreviations}
AI: artificial intelligence; API: application programming interface;
CI: confidence interval; CLI: command-line interface;
GDPR: General Data Protection Regulation; HTTP: Hypertext Transfer Protocol;
IQR: interquartile range; JSON: JavaScript Object Notation;
JWKS: JSON Web Key Set; JWT: JSON Web Token;
LLM: large language model; MCP: Model Context Protocol;
PDPL: Personal Data Protection Law; PII: personally identifiable information;
RLS: row-level security; SQL: Structured Query Language;
TLS: Transport Layer Security.

\section*{Declarations}
\subsection*{Availability of data and materials}
\textbf{Data and code availability.} The sanitized research data supporting
this study and the synthetic reference implementation code will be provided
on request from the corresponding author, subject to applicable confidentiality
and data-protection restrictions. Production database records, credentials,
and deployment-specific configuration will not be shared.

\section*{License and citation}
Copyright \textcopyright~2026 the authors. This preprint is licensed under the Creative Commons
Attribution 4.0 International License (CC BY 4.0), available at
\url{https://creativecommons.org/licenses/by/4.0/}. Sharing or adapting this
work requires the attribution specified by that license. In a scholarly
publication, provide that attribution by citing this article. The runnable
reference implementation is separately licensed under Apache-2.0; authors of
scholarly work that uses the software are requested to cite this article using
the repository's \texttt{CITATION.cff} metadata.

\section*{Competing interests}
Mirza Samad Ahmed Baig, Asher Ali, and Muhammad Hamzah Siddiqui are affiliated
with Fandaqah, the organization operating the evaluated system.

\end{document}